\pdfoutput=1

\documentclass[10pt,journal,comsoc]{IEEEtran}
\usepackage[T1]{fontenc}
\usepackage[utf8x]{inputenc}

\usepackage{ifpdf}

\ifCLASSOPTIONcompsoc
  \usepackage[nocompress]{cite}
\else
  \usepackage{cite}
\fi

\ifCLASSINFOpdf
  \usepackage[pdftex]{graphicx}
\else
\fi
\usepackage{amsmath}
\usepackage[cmintegrals]{newtxmath}
\ifCLASSOPTIONcompsoc
  \usepackage[caption=false,font=normalsize,labelfont=sf,textfont=sf]{subfig}
\else
  \usepackage[caption=false,font=footnotesize]{subfig}
\fi

\usepackage{soul}
\usepackage{color}

\usepackage{booktabs}
\usepackage{tabularx}
\usepackage{multirow}

\usepackage{pdfcomment}
\usepackage{todonotes}

\usepackage{changes}
\definechangesauthor[name={Per cusse}, color=orange]{per}
\setremarkmarkup{(#2)}

\usepackage{ulem}
\usepackage{listings}
\usepackage{filecontents}

\usepackage{lipsum}
\usepackage{url}
\usepackage{multirow}
\usepackage{tabularx}
\usepackage{tabulary}
\usepackage{booktabs}
\usepackage{rotating}
\usepackage{algorithmic}
\usepackage{algorithm}
\usepackage{xcolor}
\usepackage{longtable}
\usepackage{acronym}
\usepackage{verbatim}
\usepackage{listings}
\usepackage{float}
\usepackage{subfloat}
\usepackage{soul}
\usepackage{foreign}

\usepackage{enumitem}
\usepackage[all=normal,floats]{savetrees}

\makeatletter
\DeclareRobustCommand\ie{%
  \UKUS@comma{i.e}%
}
\DeclareRobustCommand\eg{%
  \UKUS@comma{e.g}%
}
\makeatother

\definecolor{lightgray}{gray}{0.9}
\definecolor{Gray}{gray}{0.75}
\definecolor{Blue}{rgb}{0, 0.5, 1}
\definecolor{Red}{rgb}{1, 0, 0}
\definecolor{LightGreen}{rgb}{0.85, 0.9, 0.854}
\definecolor{DarkPeach}{rgb}{1, 0.854, 0.725}
\definecolor{White}{rgb}{1, 1, 1}

\usepackage[absolute,showboxes]{textpos}

\TPMargin{5pt}

\newcommand{\copyrightstatement}{
    \begin{textblock}{0.84}(0.08,0.93)    
         \noindent
         \centering
         \footnotesize
         To appear on a future issue of IEEE Transactions on Reliability. Citation information: DOI \url{https://doi.org/10.1109/TR.2019.2954384}
    \end{textblock}
}

\renewcommand{\tablename}{Table}

\begin{document}

\copyrightstatement

%
\title{Dependability Assessment of the Android OS through Fault Injection}
%
%
%

\author{Domenico Cotroneo, Antonio Ken Iannillo, Roberto Natella, Stefano Rosiello}

\maketitle

\begin{abstract}
The reliability of mobile devices is a challenge for vendors, since the mobile software stack has significantly grown in complexity. In this paper, we study how to assess the impact of faults on the quality of user experience in the Android mobile OS through fault injection. We first address the problem of identifying a realistic fault model for the Android OS, by providing to developers a set of lightweight and systematic guidelines for fault modeling. Then, we present an extensible fault injection tool (\emph{AndroFIT}) to apply such fault model on actual, commercial Android devices. Finally, we present a large fault injection experimentation on three Android products from major vendors, and point out several reliability issues and opportunities for improving the Android OS.
\end{abstract}


%
\IEEEpeerreviewmaketitle

\section{Introduction}
\label{sec:introduction}

Reliability is a major factor that affects the quality of the user experience of mobile devices \cite{capgemini2017world}. 
In a mobile context, the user perceives a poor quality of experience if the device experiences a freeze or reboot, or if the device becomes slow or unresponsive to user inputs (e.g., when the user is starting a call, opening the list of contacts, writing a text message, etc.) because of faults. 
For example, achieving a short launch time (i.e., under $200$ ms) for basic apps, such as the phone and camera apps, has been a key performance goal since the early design of the Android OS \cite{tanenbaum2014modern,android-web-perf-anr}. 
Ideally, the Android OS should be able to assure a good quality of experience despite the occurrence of faults, such as by gracefully isolating the faulty component without affecting the availability and responsiveness of the mobile device. 

Reliability is a challenging goal for device vendors, since the mobile software stack has significantly grown in size and complexity. The Android OS, which is the most widespread mobile OS in commerce \cite{statista_2017q2}, has become a huge and sophisticated system, as the Android Open-Source Project (AOSP) now exceeds 20 million lines of code \cite{AOSP}. 
This growing trend is reinforced by the fierce competition among device vendors, which need to release new products at a fast pace, and which heavily customize the mobile OS to differentiate their products \cite{wu2013impact,gallo2015security,iannillo2017chizpurfle}. 
Unfortunately, this complexity also increases the likelihood of faults from the components of the mobile stack, including hardware-abstraction libraries, native processes, device drivers,  etc. \cite{maji2010characterizing,qin2017empirical,guana2012stars}. The mobile OS needs to manage these faults to prevent failures of the mobile device.

In this paper, we study how to assess the impact of faults on the quality of user experience in the Android mobile OS.  
Fault injection is a widely recognized technique to analyze these reliability issues \cite{hsueh1997techniques,voas1997software,natella2016assessing}: The injected faults aim to bring the system to fail in a controlled environment, so that developers can study the chain of events that lead to the failure, and they can revise the system to prevent similar failures in the final product. 
Fault injection sounds very attractive but, regrettably, developers hardly put this approach in practice, since it comes with high costs and many ambiguities. In particular, it is difficult for them to define what faults to inject (\emph{fault model}) \cite{chillarege1996:generation-error-set,duraes2006:emulationswfaults,natella2018analyzing}, due to the large number and heterogeneity of hardware and software components in a modern mobile device. 

Our assessment methodology aims to support developers at defining, conducting, and analyzing fault injection tests in a systematic and efficient way. Towards this goal, the paper presents the following novel contributions:

\begin{itemize}[leftmargin=4mm]

\item \emph{A lightweight fault modeling approach}. The approach guides developers at defining fault models through an architectural review of the mobile OS, by combining systematic, lightweight guidelines with experts' knowledge in order to be applicable in large software systems.

\item \emph{A comprehensive fault model for the Android OS}. We derived a fault model including over 600 software and hardware-induced faults, and spanning across several areas of the Android OS, including connectivity management, telephony, multimedia, sensors, and basic OS components. 

\item \emph{An Android fault injection tool ({AndroFIT})}. The tool enables the implementation of the fault model through reusable fault injection plugins, and an extensible architecture for targeting resources and service interfaces in the Android OS. Moreover, the tool has been designed to run experiments on actual Android devices (i.e., not in an emulated environment) and does not rely on the availability of Android source code (i.e., it is applicable to proprietary, closed-source versions of the Android OS).

\item \emph{A large experimental fault injection campaign}. We assessed three Android devices from major commercial vendors (Huawei, HTC, Samsung), pointing out several opportunities for improving reliability, including the lack of error handlers, weak protocol and format parsers, and UIs prone to stalls and performance degradation.

\end{itemize}

This paper is structured as follows. In Section~\ref{sec:fault_inj_methodology}, we give an overview on fault injection and on the Android OS. Section~\ref{sec:fault_modeling_android} derives a fault model for the Android OS. In Section~\ref{sec:androfit_tool}, we introduce our fault injection tool. Section~\ref{sec:experiments} presents experimental results, and discusses reliability issues and mitigation strategies.  Section~\ref{sec:related_work} discusses related work.  Section~\ref{sec:conclusion} concludes the paper.

\section{Background on the Android OS and overview of the fault injection methodology}
\label{sec:fault_inj_methodology}

The goal of our dependability assessment is to evaluate the impact of faults on the quality of user experience in Android devices. Therefore, our methodology injects faults in individual components of the mobile stack, and analyzes the responsiveness and availability of the system \emph{as a whole} in the presence of the injected fault.

A key part of the methodology is on defining a \emph{fault model}, that is, a specification of what faults to inject in the Android stack. This aspect can be tricky, as faults can originate from any hardware and software component of the mobile stack. 
Since the Android OS is a large, complex software system, the methodology gives emphasis on injecting \emph{software faults}, which are a major cause of reliability issues in such systems \cite{chillarege1996:generation-error-set,duraes2006:emulationswfaults,natella2018analyzing}. Software faults are residual defects (i.e., bugs) that escape the software development process; in our context, software faults can originate both from the open-source Android development, and from proprietary customizations introduced by the device vendors and its suppliers \cite{maji2010characterizing,wu2013impact,gallo2015security,iannillo2017chizpurfle}. Moreover, the methodology considers \emph{faults in the software that are induced by the hardware I/O peripherals} \cite{hsueh1997techniques,gunawi2011fate,qin2017empirical}. These faults manifest themselves in software as unavailable or corrupted I/O devices in the Linux kernel. Examples of both categories of faults include: buffer overruns and other forms of data corruption; exhaustion of memory, CPU, and other resources; synchronization errors, such as delayed or lost events; API or ABI incompatibilities; I/O and filesystem errors from device drivers and the kernel.

\begin{figure}[htb!]
  \centering 
  \includegraphics[width=\columnwidth]{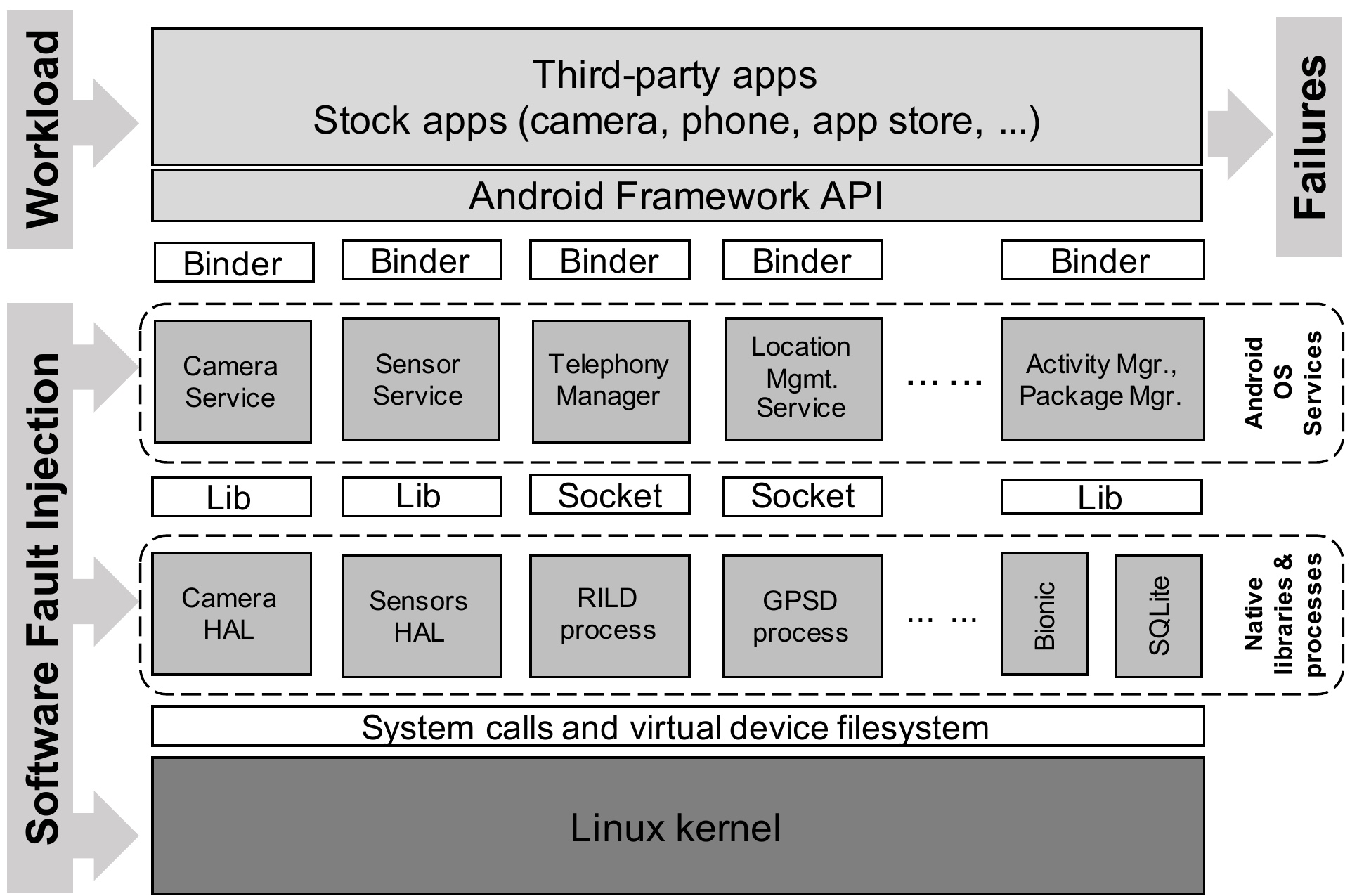}
  \caption{Architectural overview of the Android OS.}
  \label{fig:android_architecture}
\end{figure}

The methodology performs a series of fault injection experiments, where each experiment emulates a fault in one of the components at the lower levels of the Android software stack\cite{yaghmour2013embedded}, showed in \figurename{}~\ref{fig:android_architecture}. The injected components are denoted as ``\emph{fault injection targets}''. 
The uppermost layer of the Android stack is represented by the \emph{Android Framework API}, which is a collection of Java classes that provide programming abstractions to mobile apps. This API communicates with \emph{Android OS services}, which are mostly written in Java and which run on separate, privileged processes: for example, the \emph{Media Server} process runs the \emph{Camera Service}, and the \emph{System Server} runs the \emph{Activity Manager} and many other core services. Android services use hardware resources by interacting with the \emph{Native} layer, which consists of code libraries and separate processes written in C and C++. All these layers communicate both through traditional UNIX IPC mechanisms (e.g., UNIX sockets), and through the \emph{Binder}, a novel IPC mechanism based on remote procedure calls (RPCs). Finally, the Linux kernel provides access to the hardware resources through system calls and the virtual device filesystem (i.e., virtual device files in \texttt{/dev/}). 
Examples of fault injection targets related to the camera device are: the \emph{Camera HAL} (\emph{hardware abstraction layer}), which is a (typically proprietary) library that handles the camera peripheral and performs image pre-processing; the \emph{Camera Service}, which exposes an abstract, callback-oriented interface for mobile apps; and the low-level device driver for the camera in the Linux kernel.

During an experiment, the Android device is exercised with a \emph{workload} that runs a set of mobile apps, and that emulates user inputs on these apps. The workload is aimed at exercising the system (e.g., by stimulating the camera app), in order to let it experience the injected fault (e.g., in the Camera Service or HAL). Moreover, the workload exercises other areas of the system (such as networking, storage, sensors, etc.), in order to assess whether the faults escalates into a failure in any of these areas. 
Potential failures include crashes and stalls of stock apps by the device vendor and of native OS processes; 
and unrecovered errors of Android OS services (such as the Camera Service, the Telephony Registry, etc.) that are denoted by high-severity error messages from the Android OS.

The following of this paper provides more information on how this general methodology can be applied in practice, by focusing on the issue of defining a fault model for the Android OS. Then, the paper delves into the practical implementation of the methodology through automated tools.

\section{Fault modeling in the Android OS}
\label{sec:fault_modeling_android}

The Android OS is a complex software system, with many heterogeneous components. This heterogeneity makes fault modeling difficult, since components can exhibit different types of faults. The fault model for a component depends on several factors, including the resources that are used by the component, and the amount and type of services that are provided by the component. For example, if the component adopts dynamic resource allocation or concurrency (e.g., it is multi-threaded or event-driven), then it can be prone respectively to resource leaks and to synchronization faults. Therefore, in this paper, we introduce the \emph{Service Interfaces and Resources} (SIR) approach to define a realistic fault model for the Android OS in a practical, yet systematic way. 

Our fault modeling approach is founded on the observation that Android is a service-oriented system, with a well-defined software architecture (\figurename{}~\ref{fig:android_architecture}). In Android, software components have two fundamental roles: they are providers of \emph{services} that are consumed through well-defined interfaces by remote procedure calls, libraries, sockets, system calls, and other communication mechanisms; and they are managers and users of \emph{resources}, such as memory, threads, processes, communication channels, and hardware devices. The interactions between a component and the rest of the system (other OS components, apps, and physical phone) must necessarily pass through these service interfaces and resources.

Therefore, the first step of the SIR approach is the \textbf{analysis of the software architecture}. We define a list of all the service interfaces exposed by each OS component, along with its resources. This information can be obtained from the documentation of the OS, from the inspection of the source code when available (e.g., open-source components) and of the binary code (e.g., by inspecting the dependencies of a binary executable), and from queries to the OS kernel about resource utilization at run-time. It is worth noting that fault modeling does not require a full understanding of components' internals, but only the identification of interfaces and resources.

The second step of the SIR approach is to \textbf{define failure modes} for the service interfaces and resources that were identified by the architectural analysis. These failure modes will be injected for assessing the dependability of the Android OS. 
The SIR defines the failure modes using lightweight guidelines and checklists, to allow developers to systematically identify which of a pre-defined failure types apply to each component and to each interface/resource. We defined these guidelines and failure types on the basis of the scientific literature on dependable computing \cite{barton1990fault,cristian1991understanding,siewiorek1993development,mukherjee1997measuring} and on our previous work on modeling the effects of software faults \cite{natella2018analyzing}.

\begin{figure}[!thb]
  \centering
  
  \includegraphics[width=\columnwidth]{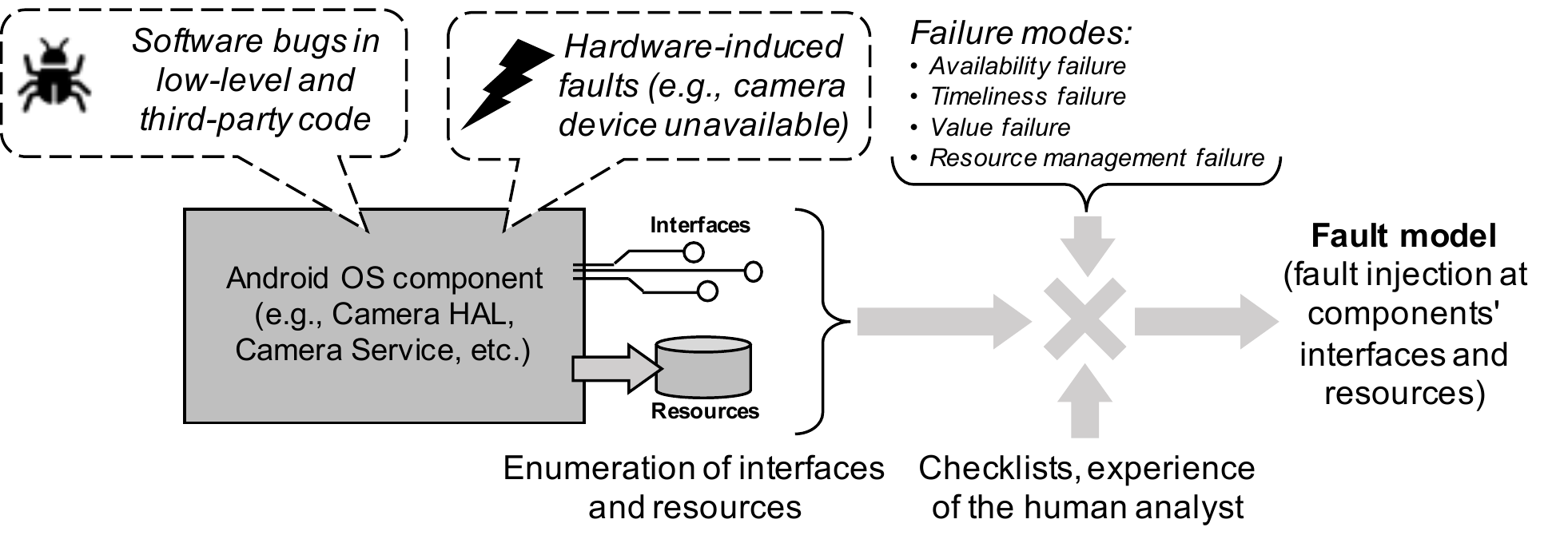}

  \vspace{-5pt}

  \caption{Relationship between faults, failure modes, and fault injection.}
  \label{fig:sir_component}

\end{figure}

The \emph{failure modes} are the potential outcomes of faults that can occur inside the component, and that can affect the rest of the system through components' interfaces and resources (\figurename{}~\ref{fig:sir_component}). 
The fault modeling approach focuses on failure modes, rather than their root cause (i.e., the originating faults), since they are both more efficient to inject through automated tools\footnote{This form of injection is in some cases referred to as \emph{error injection} or \emph{failure injection}. According to the terminology of Avizienis et al. \cite{avizienis2004:basicconcepts}, the failures of a component can be considered as faults from the perspective of the system that integrates the component. Therefore, it is commonplace to still refer to this form of injection as \emph{fault injection}.}, and easier to reason about for a human analyst:

\begin{itemize}[leftmargin=4mm]

\item Injecting faults inside a component (by mutating its source or binary code \cite{duraes2006:emulationswfaults,van2016hsfi,cotroneo2012experimental,cotroneo2018faultprog}) can be inefficient, since component-internal injections are often \emph{dormant}, i.e., the injections do not change the behavior of the component as perceived by the rest of the system; therefore, these dormant injections cannot test fault tolerance, and result in a waste of experimental time \cite{hsueh1997techniques,voas1997software,natella2016assessing}. Instead, the injection of failure modes in component's interfaces and resources is more efficient, as it directly injects the intended effects of component-internal faults \cite{tsai1999stress,jarboui2002analysis,moraes2006:interfacefaults,lanzaro2014empirical,natella2018analyzing}.

\item Moreover, from the perspective of a human analyst, it is easier to define the fault model by starting from the analysis of component's interfaces and resources, which are simpler to enumerate than the (very large) set of software faults that can possibly occur inside the component's internals (such as the examples of software faults mentioned in Section~\ref{sec:fault_inj_methodology}). 

\end{itemize}

The SIR approach identifies which failure modes are applicable to each interface and resource, and valid combinations are added to the fault model for the component. Then, we apply fault injection at the interfaces of the component, and assess the robust behavior of the rest of the system. 
We consider four types of failure modes, which were derived from general failure classifications by previous studies, including Barton et al. \cite{barton1990fault}, Cristian \cite{cristian1991understanding} and Siewiorek et al. \cite{siewiorek1993development}; the similarities between these failure classifications were compared by Mukherjee and Siewiorek \cite{mukherjee1997measuring}. The component failure modes include:

\begin{itemize}[leftmargin=4mm]

\item The component refuses to provide a service to its user, by returning an exception or error for the request (\textbf{availability failure});

\item The component provides a service response only after a long delay, or it does not return at all while its user waits indefinitely (\textbf{timeliness failure})

\item The component produces a wrong service response, by returning incorrect data to its user (\textbf{output value failure});

\item The component saturates, exhausts, or disables the resources that it uses or manages (\textbf{resource management failure}).

\end{itemize}

The checklists for assigning these failure modes are based on previous experience of the research community on fault injection \cite{duraes2006:emulationswfaults, chillarege1996:generation-error-set,voas1997:predicting,marinescu2009lfi,gunawi2011fate,lanzaro2014empirical}. 
The first checklist focuses on components' \textbf{services}. The checklist includes questions to identify which of the four generic failure modes can occur for each service interface. A fault is added to the final fault model if the scenario is plausible according to the checklist:

\begin{enumerate}[leftmargin=4mm]

\item Does the service interface declare exceptions, or erroneous return codes? If yes, add an \emph{availability failure} for the service.

\item Can the service lose a request (e.g., due to a service queue overflow) or a reply (e.g., an asynchronous method call is not followed by a response callback), without performing any operation? This possibility should be considered when the component is multi-threaded or event-driven. If yes, add a \emph{timeliness failure} for the service.

\item Can the service experience a long delay? This possibility should be considered if the component processes large amounts of data or performs high-volume I/O activity, which may lead to performance bottlenecks. If yes, add a \emph{timeliness failure} for the service.

\item Can the service return a result (e.g., a numerical computation or a data structure) that may be incorrect due to a bug? This possibility should be considered if the service implements non-trivial algorithms or manages complex data structures (e.g., services that make use of concurrency, dynamic resource allocation, linked data structures, etc.) \cite{lanzaro2014empirical,natella2018analyzing}. If yes, add an \emph{output value failure} for the service.

\end{enumerate}

In a similar way, the second checklist focuses on components' \textbf{resources}:

\begin{enumerate}[leftmargin=4mm]

\item Can the process or thread that hosts the component experience a crash (i.e., killed by the OS), or terminate prematurely, or be stalled (e.g., because of a deadlock), before replying? This possibility should be considered when the component is relatively large (several thousands of lines of code) and includes native code. If yes, add a \emph{resource management failure} for the use of processes or threads.

\item Is the resource protected by permissions, and can it become inaccessible due to lack of permission? For example, this is the case of inter-process shared resources in UNIX systems. If yes, add a \emph{resource management failure} for the resource.

\item Can the component leak the resource (e.g., memory and file descriptors that are frequently allocated/deallocated), thus preventing further allocations of the resource? If yes, add a \emph{resource management failure} for the resource.

\item Does the component allocate new processes or threads? These may terminate prematurely, or the component may hit hard system limits when allocating them (e.g., \emph{ulimit} in UNIX systems). If yes, add a \emph{resource management failure} for the use of processes or threads.

\item Does the component manage persistent files (e.g., a database file or a configuration file) that may be corrupted when reading or writing it? If yes, add a \emph{resource management failure} for the corruption of the file.

\end{enumerate}

Finally, for every item in the fault model, the SIR approach adds information on the \textbf{persistence} of the faults, which indicates the behavior of the injected fault over time, i.e., whether it is \emph{permanent} (the fault persists for a long period of time) or \emph{transient} (the fault occurs only in a specific moment of the execution). 
The fault is flagged as permanent if the fault's effects are persistent unless explicitly recovered or cleaned (for example, a resource leak or a crash); or as transient, if the fault is triggered by a rare environmental condition (such as an external event). 
In the case that a fault can exhibit more than one kind of persistence, we introduce multiple entries for the fault in the fault model.

The SIR approach provides generic guidance for test engineers, but it still leaves room for the human judgment, as it is their call to decide whether a service is ``complex'' or a condition is ``rare'' when applying the checklists. During our work on the fault model for the Android OS, we cooperated with the test engineers of a major Android vendor, by asking them if a fault could be plausible according to their knowledge and experience with the Android OS. Framing the discussion in these terms helped us to iteratively improve the fault model, and to make it accepted by the stakeholders as realistic.


\subsection{The Android OS fault model}
\label{subsec:android_fault_model}

We applied fault modeling throughout components at the lower layers of the Android OS. We targeted these components as \emph{fault injection targets} (FIT), since they are subject to customizations both from the device vendor and its third-party suppliers, which tend to introduce bugs \cite{wu2013impact,iannillo2017chizpurfle}. Moreover, these components tend to be bug-prone since they deal with concurrency, resource management and the hardware. 
In total, we analyzed 27 components of Android, including Android OS Services, native components, and device drivers \cite{yaghmour2013embedded}. The components span across several areas of the Android OS, including volume and connectivity management, telephony, multimedia, sensors, and basic components of the Android OS such as the System Server, the Zygote, etc..

We first performed an architectural review of these components, to identify their service interfaces and resources. The service interfaces include Binder services, services over UNIX sockets, library APIs, and device drivers. The resource types include processes and threads, memory, ordinary files, device files, UNIX sockets and pipes, and Binder objects. 
Then, we defined a comprehensive list of failure modes for these components of the Android OS. 
In total, we identified 684 failure modes across these components, by applying the checklists for every API function, communication channel or resource of the components, and by considering the nature of the components' API (e.g., presence of return values to raise errors, output values that could be corrupted, synchronous/asynchronous nature of the API, etc.). 
Since the fault model is large, we reported the full list in a separate document, available as open data on the web (\url{https://figshare.com/s/406f05f915ec517e675c}).

\subsection{An example from the phone subsystem}
\label{subsec:phone_fit_example}

\begin{figure}[htb!]
  \centering 
  
  \vspace{-5pt}
  
  \includegraphics[width=0.75\columnwidth]{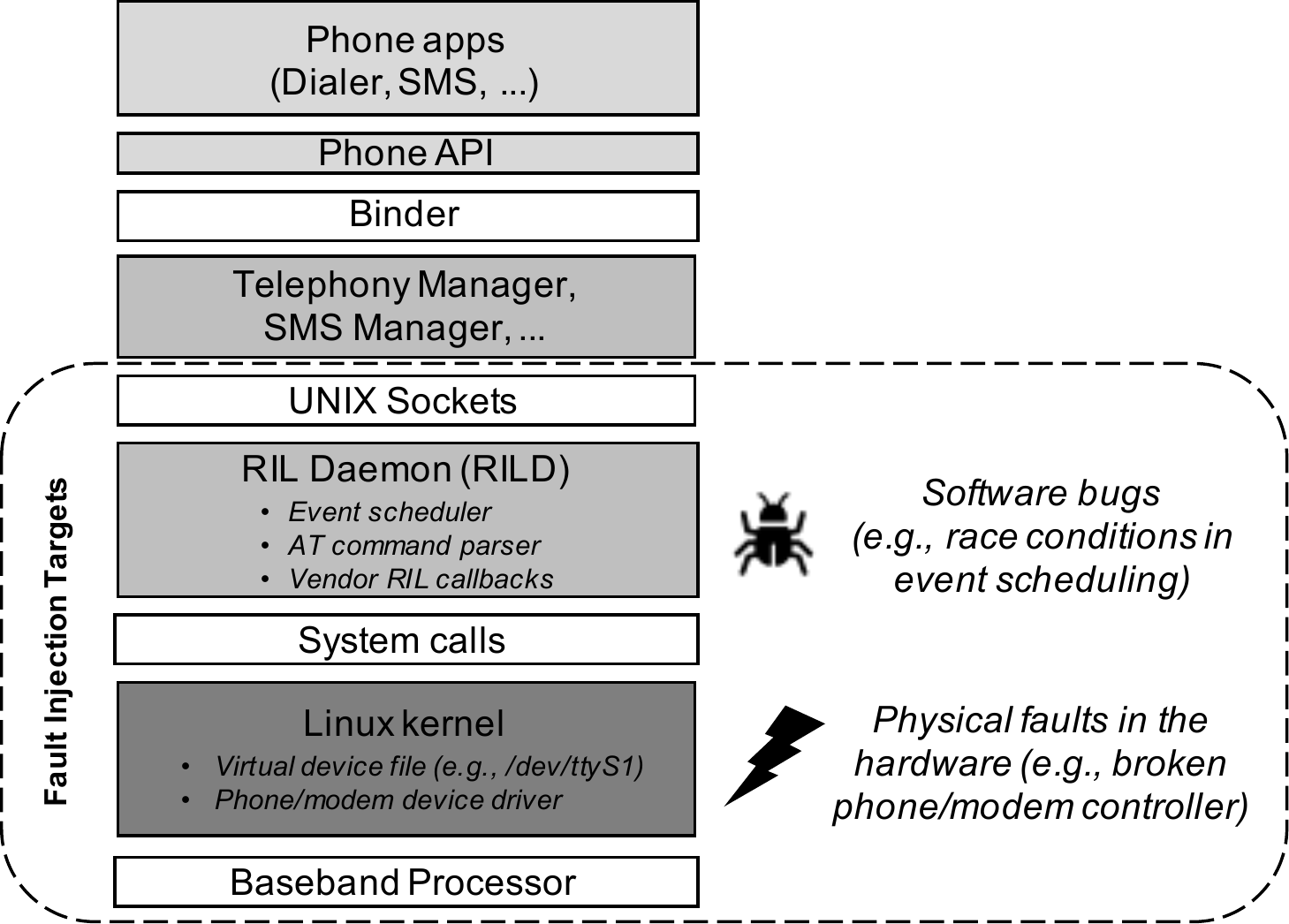}
  
  \vspace{-5pt}
  
  \caption{Examples of fault injection targets.}
  \label{fig:phone_fit_example}
\end{figure}

As an example, we consider the phone subsystem as a running example for this paper. 
The phone subsystem (\figurename{}~\ref{fig:phone_fit_example}) includes the following components:

\begin{itemize}[leftmargin=4mm]

\item \textbf{RILD}: a system process that embeds a proprietary, vendor-specific RIL (Radio Interface Layer) library and the Event Scheduler, which dispatches the events from the baseband processor, and the commands from the upper layer;

\item \textbf{Baseband Driver and Processor}: Baseband Driver exposes a device file (e.g., \texttt{/dev/ttyS1} or \texttt{/dev/ttyUSB1}) to send/receive commands and events to/from the Baseband Processor, which controls the physical communication with the network.

\end{itemize}

The RILD and the Baseband Driver represent potential fault injection targets. The RILD provides services over a UNIX socket, and consumes services of the Baseband Driver through system calls on a device file. To apply the SIR methodology, we obtained the list of all the service interfaces and resources for the RILD and Baseband Driver, by looking at public information on the Android architecture, and at run-time information from the Linux kernel (e.g., using \emph{netstat} for inspecting UNIX sockets, and \emph{lsof} for file descriptors).

The RILD provides two forms of services to the upper layers of the Android stack. It receives and executes phone \emph{commands} from the phone library (e.g., on behalf of the phone stock app) to start a call, to send a message, etc.; and it sends phone \emph{events} to the phone library, such as, to notifying them that a call has been dropped. These RILD provides services by using lower-level services from the Baseband Driver, by writing commands for the modem, and by reading responses from the modem, using the AT protocol. 
The RILD resources include the RILD process and its threads; the memory used for handling and queueing commands and events; the UNIX socket for communicating with the phone library. Moreover, the Baseband Driver exposes a virtual device file resource.

The service interfaces for these components are based on socket and file primitives, such as \emph{receive}, \emph{send}, \emph{read}, and \emph{write}. They all declare erroneous return codes that can be encountered during service. Thus, we introduce availability failures for these primitives. 
Moreover, the RILD service is a multi-threaded service that could be flooded by several messages in a short amount of time, or can be affected by synchronization issues, which can cause the loss of service commands and events. Thus, we introduce timeliness failures. Similarly, we added timeliness failures to encompass potential delays that can be accumulated when handling many messages. 
Since the data handled by the RILD can be accidentally altered because of faults in the data transfer procedures (e.g., due to buffer overflows), we introduce output value failures for the services of the RILD and the Baseband Driver. 
Finally, we introduce resource management failures. Since the RILD is hosted by a specific native process, we consider the possibility that it could crash or hang. Sockets and virtual device files can be unavailable due to issues with permissions, or due to resource exhaustion caused by leaks. Similarly, memory in the RILD process can become unavailable to due memory leaks.

\section{The AndroFIT fault injection tool}
\label{sec:androfit_tool}

\begin{figure*}[htb!]
  \centering 
  \includegraphics[width=0.8\textwidth]{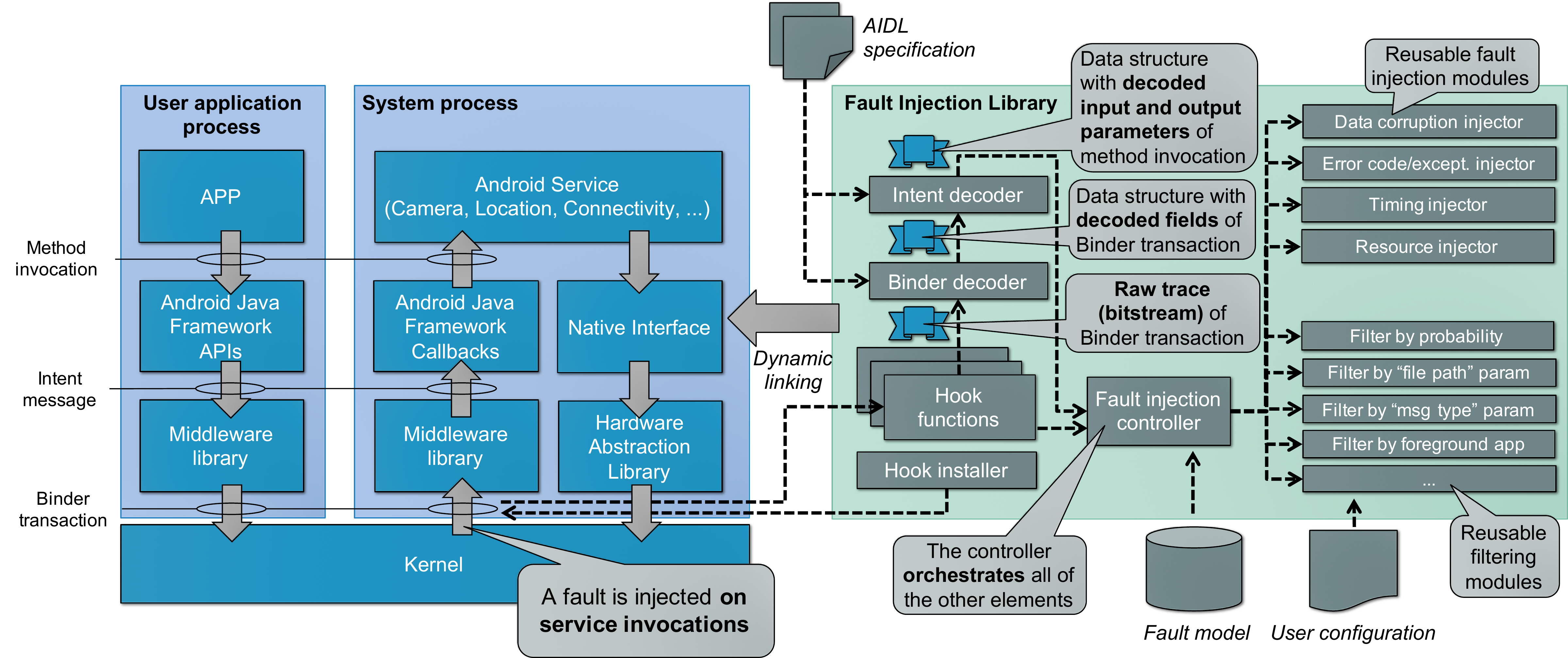}

  \vspace{-5pt}

  \caption{Overview of the AndroFIT tool.}
  \label{fig:androfit}

  \vspace{-10pt}

\end{figure*}

We developed \emph{AndroFIT} (\emph{Android Fault Injection Tool}) as a generic and flexible fault injection tool, which can be configured to inject the failure modes identified by the SIR approach across all of the several service interfaces and resources in the Android OS.

\subsection{AndroFIT design}
\label{subsec:androfit_arch}

We designed AndroFIT to avoid modifications of the source code of the Android OS, thus avoiding the time-cost for rebuilding the Android code base, and enabling fault injection in commercial devices, for which the source code is unavailable. 
Therefore, we leverage UNIX mechanisms for dynamic code loading and for interposing custom functions (\emph{hooks}) in function calls inside the fault injection target. 
The key part of AndroFIT is a \emph{fault injection library}, which is dynamic shared library linked to the process in which the fault injection target runs (such as, the System Server, the Media Server, etc.). The fault injection library includes the following components.

\vspace{2pt}
\noindent
\textbf{Hook installer}: The hook installer is executed immediately when the fault injection library is dynamically loaded into the system process, using the \emph{ptrace} system call. The hook installer modifies the entry points of the component to be injected, such as the functions invoked when Binder transactions are received. In particular, the hook installer modifies the binary instructions and pointers that are used to invoke the component to be injected. These instructions and pointers are replaced with instructions or pointers to hook functions inside the fault injection library. According to the fault modeling method, the following functions should be hooked to cover the components of the Android OS:

\begin{itemize}[leftmargin=4mm]

\item \emph{Kernel-level interfaces}: they include library functions and system calls (\emph{read}, \emph{write}) that are used to access to the hardware device through a virtual device file (such as \texttt{/dev/video0}, \texttt{/dev/smd0}, etc.). The hook functions can simulate faults of device drivers and filesystems. 
We do not consider generic kernel bugs (e.g., scheduling and memory management), but focus on these modules, as they are developed outside the open-source Android project by third parties (e.g., device vendors or their suppliers) and are the kernel components most prone to failures according to previous research studies \cite{chou2001empirical,ganapathi2006windows,palix2011faults}.

\item \emph{IPC interfaces}: they include library functions and system calls (\emph{sendmsg}, \emph{recvmsg}, ...) for IPC between Android components, such as sockets, pipes, message queues, etc.. The hook functions can simulate faults on IPC-based components, such as the RILD and Zygote process that communicate with the upper layers through UNIX sockets (e.g., \texttt{/dev/socket/rild}, \texttt{/dev/socket/zygote}).

\item \emph{Binder interfaces}: they include library functions and system calls (in particular, \emph{ioctl})  to access to the Binder driver, using the virtual file \texttt{/dev/binder}. As discussed later, the Binder driver is a low-level carrier for RPC calls between Android components. Therefore, the tool also provides decoders to decode structure data (e.g., input objects to RPCs).

\item \emph{HAL APIs}: they include library functions (such as the Camera HAL API, the Sensors HAL API, etc.) that are typically accessed by Android components (based on well-defined API specification) through function pointers in shared data structures. The hooks replace the pointers to the original functions. 

\item \emph{Resource management APIs}: they include library functions and system calls to dynamically allocate and retrieve OS resources (e.g., \emph{malloc} and \emph{free} for memory, \emph{open} and \emph{close} for files, \emph{fork} and \emph{clone} for processes, \emph{msgget} and \emph{shmget} for UNIX IPC resources, etc.). 
\end{itemize}

\vspace{2pt}
\noindent
\textbf{Hook functions}: The hook functions are invoked at every invocation of the target component to be injected. Upon invocation, the hook functions trigger the fault injection controller and the other elements of the fault injection library. These elements may, or may not (depending on the fault model and on the user configuration) modify the contents or the timing of the target component invocation. Then, the hook functions return the execution of the original functions.

\vspace{2pt}
\noindent
\textbf{Binder decoder}: The Binder decoder analyzes the raw bitstream transmitted when an Android component is invoked. The Binder decoder reconstructs the original high-level Binder transaction from the low-level bitstream. The result is a data structure containing the parts of the Binder transaction, such as the message type, size, and its individual fields (such as integers, booleans, strings, etc.). This data structure allows the fault injection controller to easily access and modify the Binder transaction. The decoding process is fully automated, using the AIDL specification of the Android component interface to learn about which are the parts of a transaction.

\vspace{2pt}
\noindent
\textbf{Intent decoder}: In a similar way to the Binder decoder, the Intent decoder analyzes the raw bitstream to reconstruct \emph{Intents} (i.e., a high-level IPC abstraction used by Android apps, based on Binder). The result is a data structure with the parts of the Intent, such as the action, the category, and the class. This data structure allows the fault injection controller to easily access and modify the Intent. The decoding process is fully automated using the AIDL specification.

\vspace{2pt}
\noindent
\textbf{Fault Injection Controller}: This element orchestrates all the other elements (hooking, decoders, injection modules, filtering modules, monitoring). The Fault Injection Controller takes in input the fault model. For each entry in the fault model, the Fault Injection Controller automatically performs a fault injection according to target service/resource, failure mode, and timing described in the fault model. It first inserts a hook function for the injection point, which calls the controller when the injection point is invoked at run-time. In turn, the controller invokes the \emph{filtering modules} (discussed later) to decide whether to inject in the current invocation. 

\vspace{2pt}
\noindent
\textbf{Fault Injection Modules}: The Fault Injection Controller uses fault injection modules for modifying a service invocation or resource access. The fault injection modules are reusable for different injections across the software stack (for example, the same injection module can be used for injecting faults on the Binder for different services, such as the Camera Service and the Connectivity Manager). These modules include:

\begin{itemize}[leftmargin=4mm]

\item	\emph{API errors/exceptions}: the fault injection module forces the interaction to terminate, by returning an error code or exception;

\item \emph{Timing}: the fault injection module stalls the interaction for a period of time, or indefinitely;

\item	\emph{Data corruption}: the fault injection module modifies the input or output values of the interaction;

\item	\emph{Resource unavailability}: the fault injection module forces resource unavailability when the component uses an API for allocating or accessing a resource;

\end{itemize}

We configure error codes and exceptions according to error codes and exceptions found in AOSP for the target service function. Moreover, we apply data corruptions according to corruption patterns reported by previous studies \cite{devale1999ballista, iannillo2017chizpurfle}. The corruptions include: for categorical types (such as ``msgType''), we replace the original value with another possible value for the type; for numerical types (such as integers and floats), we replace the original value with boundary or undefined values (such as zero, off-by-one value, off-by-offset value, negative value, max/min value, randomly-selected values); for strings (such as camera parameters), we return NULL objects, replace substrings, or truncate the string; for bitmaps and raw data and metadata (such as the data of a camera image), we perform bit-flipping; for structured types (such as camera capture request), we corrupt individual fields in the structure (such as integer and categorical fields) according to their type.

\vspace{2pt}
\noindent
\textbf{Filtering Modules}: The filtering modules control when and how often faults should be injected during a test. When invoked, a filtering module checks whether a specific condition (configured by the test engineer) is satisfied. Every time that the hook function is triggered, the fault injection controller queries the filtering modules, and if all the conditions checked by these modules are satisfied, then a fault is injected in the current invocation of the target component. The fault injection framework can enable or disable separately each filtering module; filtering modules can be reused across different faults in the fault model; more filtering modules can be added by the test engineer. The filtering criteria include: \emph{probability} (i.e., injection in a randomly-selected subset of invocations, such as 10\% for transient faults, and 100\% for permanent faults); \emph{file path} (i.e., only when a parameter representing a ``file path'' contains an user-defined sub-string, such as a virtual device name); \emph{message type} (i.e., injection when a byte stream contains a user-defined pattern, such as a specific ``command'' tag sent through a socket); \emph{foreground app} (i.e., injection when the workload is interacting with a specific app). 
When configuring probabilities for fault injection, the probability is typically set higher than the actual probability of the transient faults that the injections are meant to emulate. However, a higher probability is needed since fault injection aims to accelerate the occurrence of faults (injecting the faults with their actual probability would lead to exceedingly long experiments). Therefore, we set this probability to 10\% for transient fault injections, such that the injections occur only one or few times during an experiment and are far enough over time to avoid overlaps; and yet, the probability is high enough to have at least one injection in all of the experiments (thus, avoiding the need to repeat some of them).

\vspace{2pt}

The possibility of side effects is an important concern in the design of our fault injection tool. Since the proposed SIR approach injects faults on service interfaces and resources, it comes with the added benefit that faults can be introduced by means of function call interposition (i.e., when a service or resource is accessed by means of a function call), which can be implemented with very low overhead. The only moment in which the target component is paused for a very short period of time (few ms) is when a hook function is introduced in the target by the hook installer, using the \emph{ptrace} debugging mechanism. From then on, the invocation of the hook function is performed at full CPU speed, since hardware/software debugging mechanisms are not involved anymore (for example, the hook function is invoked by following a modified function pointer, as in the case of the original function). Moreover, the operations performed by the hook functions are very limited. The hook function performs lightweight checks of the parameters of the original function, in order to determine whether to inject faults, and immediately returns the control flow to the original function if this is not the case. To inject faults, the hook function simply performs modifications to these parameters before or after that the original function is invoked. Therefore, the hook function introduces a very low latency to the execution of the call. To assure that the tool avoids side effects on the behavior of the system, we performed fault-free tests, by installing hook functions but without actually injecting the faults, and measured the additional latency introduced by the hook functions. In all fault-free tests, the function call latency introduced by the hook functions was negligible, as it was always below 1 ms, and we did not observe any failure (e.g., no app crashes or stalls).

We refer the interested reader to the PhD thesis of one of the authors \cite{kenphdthesis} for more technical information about how we implemented the AndroFIT tool.

\subsection{Test automation}

The AndroFIT tool suite includes a program, the \emph{experiment launcher}, to automate a fault injection campaign. 
The main input of the experiment launcher is a campaign configuration file that describes all the experiments, along with the desired number of repetitions of each experiment. Each entry of the campaign file configures AndroFIT to inject a failure mode (unavailability, timeliness, ...) on a specific Android service interface or resource. 
We also provide in a PhD thesis \cite{kenphdthesis} information about the automation of the experiments. Moreover, we designed AndroFIT to also provide a command-line interface (CLI) to guide the user at configuring the tool. For example, the tool provides feedback about which components and which services and resources can be injected by the tool, and raises an explanatory error message if these parameters were omitted or were entered incorrectly.

For every experiment, AndroFIT starts by generating a workload for the device, by using the \emph{monkey} tool \cite{android-monkey} to emulate user inputs across several apps and Android subsystems (e.g., switching between apps, navigating the forms of an app, etc.). 
The workload exercises popular mobile apps \cite{cotroneo2016software}, including \emph{com.tencent.mm}, \emph{com.sina.weibo}, \emph{com.qiyi.video}, \emph{com.youku.phone}, \emph{com.taobao.taobao}, \emph{com.tencent.mobileqq}, \emph{com.baidu.searchbox}, \emph{com.baidu.BaiduMap}, \emph{com.UCMobile}, and \emph{com.moji.mjweather}. 
The tool first runs the workload for a warm-up period (30 seconds). Then, while the workload is still running, the tool performs an injection. AndroFIT also performs a user-configurable action (\emph{trigger}) to stimulate the injected fault. For example, AndroFIT provides triggers for stimulating the camera (e.g., by opening the stock camera app and taking a picture), the phone (e.g., by dialing and calling a mobile phone number), and the sensors (e.g., opening and using an app to get data from sensors). The workload continues to execute until the end of the experiment. Finally, the device is rebooted, in order to have a clean device and the same initial conditions for the next experiment.

At the end of every experiment, the launcher collects log files from the device (e.g., using the \emph{logcat} tool). These logs are later analyzed to assess whether the injection has been performed, and what are the consequences of the injection on the Android OS. The potential test outcomes, and the criteria used to identify them, are:

\begin{itemize}[leftmargin=4mm]

\item \textbf{Crash}: a system process or an app crashed, and the system logs a message reporting a ``FATAL EXCEPTION'';

\item \textbf{ANR}: an app becomes stalled, and the system generates a log message that reports an ANR condition (i.e., Application Not Responding);

\item \textbf{Fatal}: the Android OS experiences a critical error, and the system generates log messages with a high-severity level, i.e., tagged either with \emph{assert} (``A'') or \emph{fatal} (``F'');

\item \textbf{No failure}: the Android OS does not exhibit any of the effects listed above.

\end{itemize}

We remark that other test outcomes are also possible under some of the injections (e.g., the ones corrupting data), such as incorrect results and GUI glitches, but these cases are beyond the scope of the experimentation.

\section{Experimental evaluation}
\label{sec:experiments}

We apply the fault injection methodology on actual Android devices, and present the results of their dependability evaluation. This evaluation serves both for demonstrating the practical application of the methodology, and for pointing out reliability issues found in commercial versions of the Android OS. 
We performed fault injection experiments on three devices from major Android vendors: the \emph{Huawei P8}, the \emph{Samsung S6 Edge}, and the \emph{HTC M9 One}. The experiments ran on the actual devices, by getting root permissions and by installing the AndroFIT tool on the device. 
We updated the Android OS on the devices to the latest version available from the vendor, thus running Android 6 on the Huawei P8, and Android 7 on the Samsung S6 Edge and on the HTC M9 One.

In a first round of experiments, we injected faults in components across the following three vertical subsystems of the Android OS:

\begin{itemize}[leftmargin=4mm]

\item \textbf{Phone}: Faults were injected in the RILD process, by targeting the socket interface between the RILD and the Android framework, and in the system call interface between the RILD and the device driver for the phone and modem.

\item \textbf{Camera}: Faults were injected in the MediaServer process, by targeting the Camera Service and the Camera HAL, and in the system call interface between the MediaServer and the device driver for the camera.

\item \textbf{Sensors}: Faults were injected in the Sensors Service, in the Sensors HAL, and in the system call interface of the device drivers for the sensors.

\end{itemize}

Moreover, in a second round of experiments, we injected faults in the following basic components of the Android OS:

\begin{itemize}[leftmargin=4mm]

\item \textbf{System Server}: A privileged process that runs the main services of the Android OS. Faults were injected in the Activity Manager, which is the service that handles the lifecycle of Android apps (e.g., pausing, stopping, switching, etc.), and the Package Manager, which is the service that handles information about resources and security permissions of Android apps.

\item \textbf{Surface Flinger}: A privileged process that receives window layers (``surfaces'') from multiple sources (such as apps, the notification bar, the system user interface, etc.), and combines and displays them on the screen. Faults were injected both in the Surface Flinger service and process.

\item \textbf{Native libraries}: We targeted two native libraries used by system processes in Android. We targeted \emph{Bionic}, which implements the standard C library, and \emph{SQLite}, which embeds a SQL DBMS. 
We injected in APIs for accessing the filesystem on the internal device storage (e.g., by truncating and corrupting the contents of I/O reads), and in APIs for performing SQL queries (e.g., by truncating and corrupting the results of queries).

\end{itemize}

\subsection{Overview of experimental results}

\begin{figure*}[!htbp]
       \centering
       \subfloat[Phone faults.]{\label{fig:benchmark_results_phone}\includegraphics[keepaspectratio=true, width=0.3\textwidth]{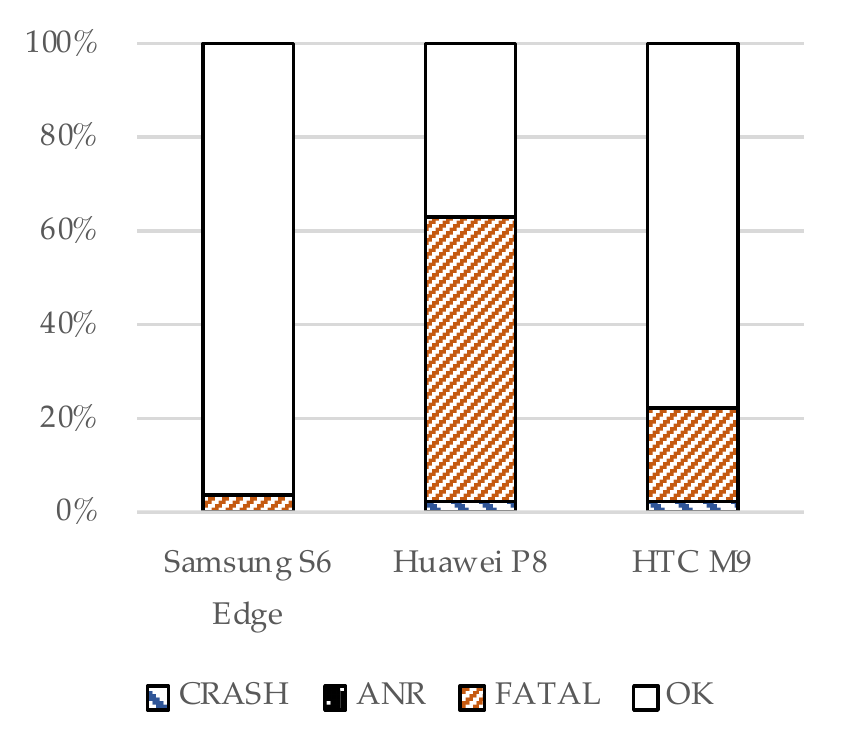}}
       \subfloat[Camera faults.]{\label{fig:benchmark_results_camera}\includegraphics[keepaspectratio=true, width=0.3\textwidth]{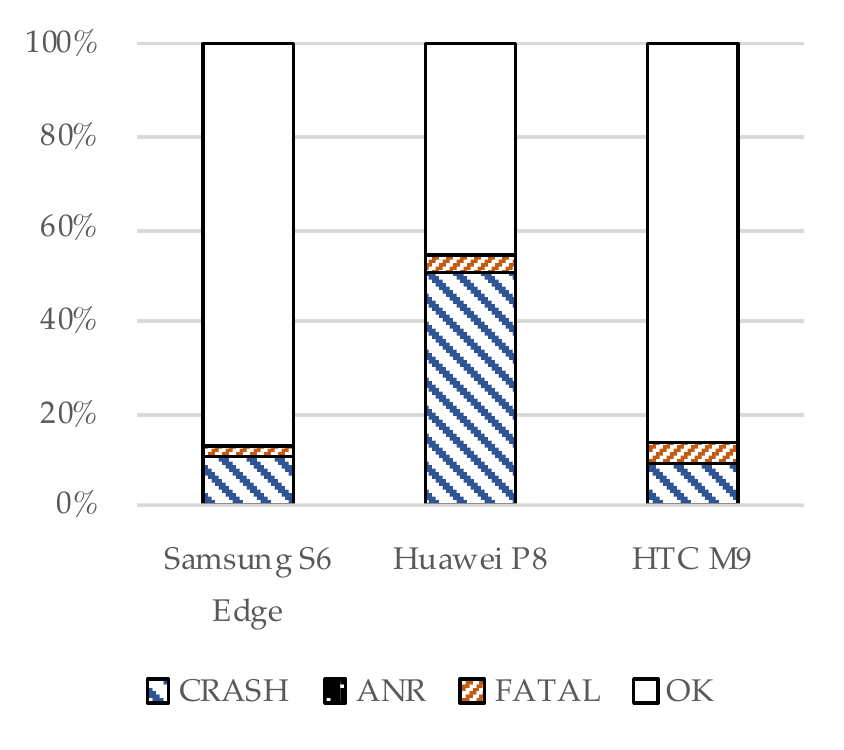}}
       \subfloat[Sensors faults.]{\label{fig:benchmark_results_sensors}\includegraphics[keepaspectratio=true, width=0.3\textwidth]{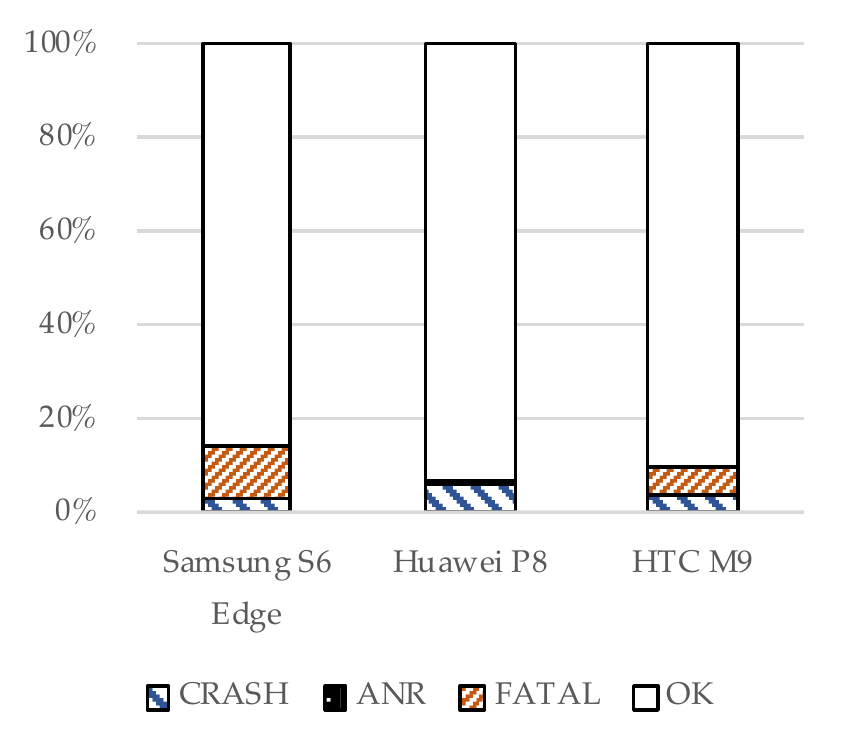}}
       
       \vspace{-5pt}
       
       \subfloat[System Server faults.]{\label{fig:benchmark_results_system_server}\includegraphics[keepaspectratio=true, width=0.3\textwidth]{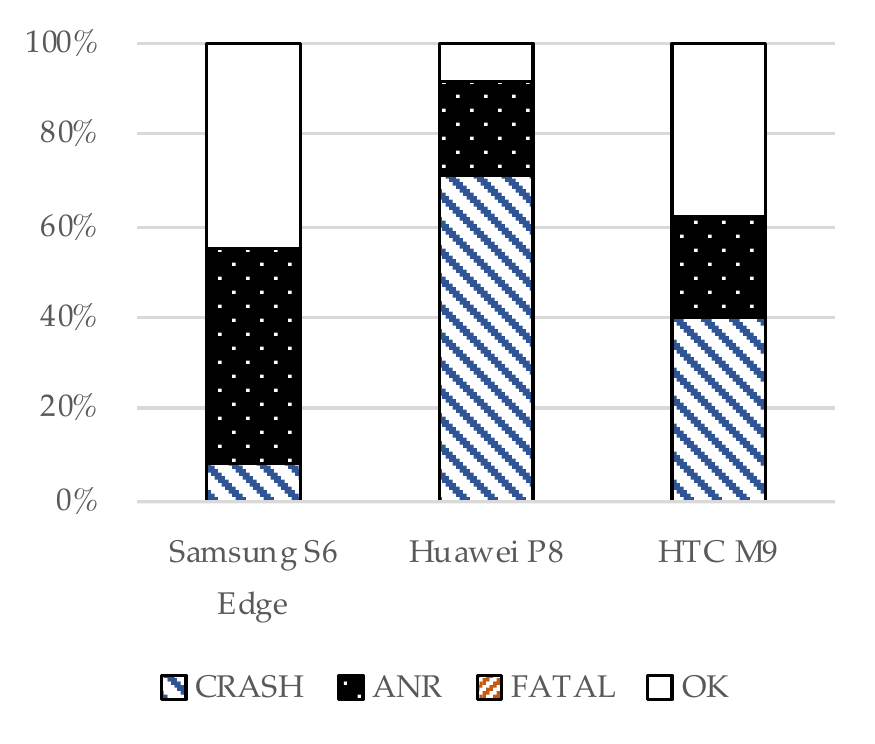}}
       \subfloat[Surface Flinger faults.]{\label{fig:benchmark_results_surface_flinger}\includegraphics[keepaspectratio=true, width=0.3\textwidth]{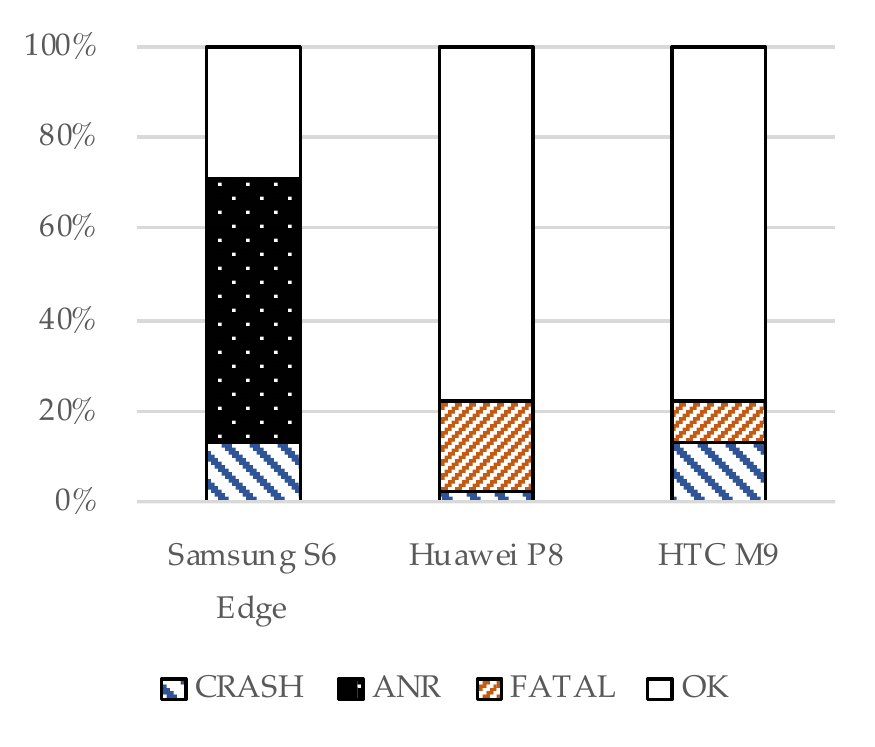}}
       \subfloat[Native library faults.]{\label{fig:benchmark_results_native_libs}\includegraphics[keepaspectratio=true, width=0.3\textwidth]{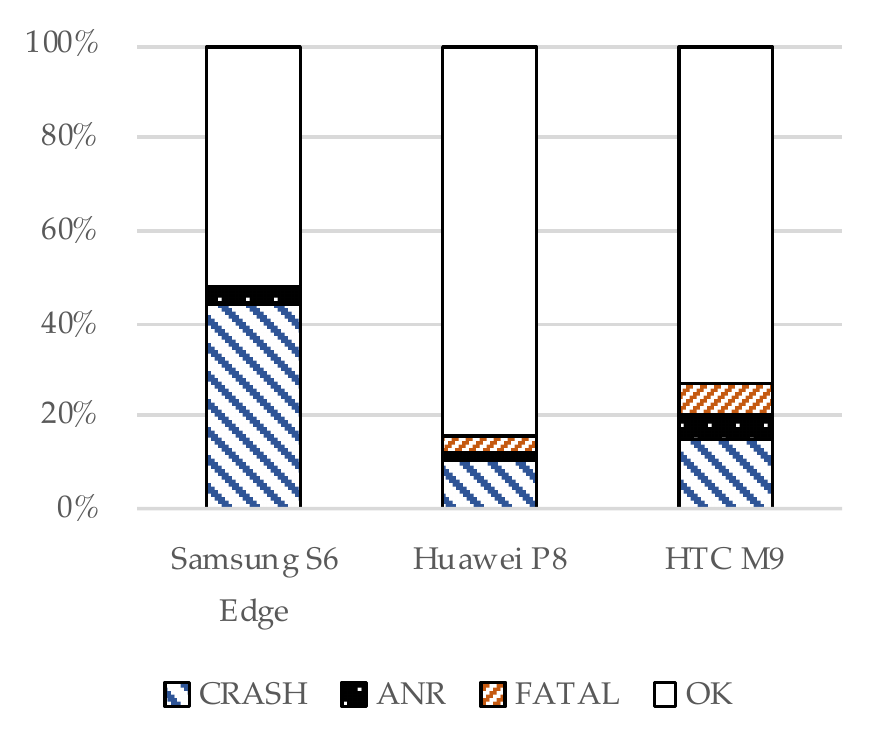}}

        \caption{Distribution of failures across different devices and different fault injection targets.}
        \label{fig:failure_distributions}
        
       \vspace{-10pt}
       
\end{figure*}

The plots in Figure~\ref{fig:failure_distributions} shows the aggregated results for fault injection across the target subsystems. The bars represent the distribution of failures, where the sub-bars are the four possible outcomes of the experiment (crash, fatal error, ANR, no failure). The same type and amount of faults were injected on every device, with three repetitions per fault. 
We performed a statistical test to assess whether the failure distributions are statistically significant and not due to chance. For each device and for each injected subsystem, we separated the data for the three repetitions of the fault injections on that device and subsystem. Then, we used the Fisher's exact test to assess the null hypothesis that the proportions of failures across the three repetitions are independent (i.e., repeating the experiments does not affect the failure distributions). In all cases, we could not reject the null hypothesis with a 95\% confidence level (all p-values greater than 0.05, omitted for the sake of brevity), as the proportions of failures were very similar across the repetitions. 
Thus, the statistical test cannot conclude that the repetitions have statistically-significant differences.

For fault injection in the phone, camera, and sensors subsystems (Figures~\ref{fig:benchmark_results_phone}, \ref{fig:benchmark_results_camera}, and \ref{fig:benchmark_results_sensors}), we performed respectively 309, 111, and 108 fault injection experiments. 
The numbers of fault injections depend on the set of service interfaces and resources that are provided by the injected subsystems, and that are actually exercised by the system under the workload. Thus, the numbers are different across subsystems since they are quite diverse in terms of service interfaces and resources. For example, to use the Camera subsystem, the mobile apps exercise a relatively smaller service interface of the CameraService API.

In the cases of the phone and camera subsystems, the Huawei device experienced the highest number of failures. Instead, with respect to faults in the sensors subsystem, the amount of failures was comparable across devices, where the Huawei device was the most robust one. These differences point out that vendor customizations can have an influence on the robustness of the Android OS against faults, thus emphasizing the importance of fault injection for testing these customizations. 
In the phone subsystem, most of the failures were ``fatal errors'', with cases of RILD process crashes for the Samsung and Huawei devices. In the camera subsystem, most of the failures were process crashes, and in particular, crashes of the camera stock application. In few cases, the camera system reported fatal errors. 
For the sensors subsystem, there were few ANR failures in the Huawei device not happened in the other devices, due to differences in the stock apps.

For fault injection on the System Server, we performed 129 experiments (Figure~\ref{fig:benchmark_results_system_server}), and we observed 71 failures for Samsung, 118 failures for Huawei, and 80 failures for HTC. The number of failures has been very high for all of the three devices. These failures (in particular, ANRs) froze the system UI and other apps (including stock apps, such as the Camera), which did not respond to the inputs of the users. In particular, these freezes have been caused by injected delays on key methods of the Activity Manager and Package Manager (such as ``bind service'', ``resolve intent'', ...) which are called during the initialization of apps (such as the Camera) and when the System UI performs special actions (such as clicking on the button for ``show all activities''). Notably, the failure rate for the Huawei device has been significantly higher, since these methods are used more in this system. For example, the Camera app was especially vulnerable for Huawei, while the Camera app of the other devices avoided the problem and was still able to start quickly, since it has less dependencies on the methods of the Activity/Package Manager.

For fault injection on the Surface Flinger, we performed 45 experiments (Figure~\ref{fig:benchmark_results_surface_flinger}), and we observed 32 failures for Samsung, 10 failures for Huawei, and 10 failures for HTC. Among the three vendors, the Samsung device had a much higher number failures. The variability is due to the fact that the vendors heavily customize the window manager, in order to show to the user visible differences between their devices and the competition. In the case of Samsung, the customizations exposed to more frequent app freezes when the Surface Flinger was slowed and returned errors. However, there were also cases in which the Huawei device exhibited a non-robust behavior in the case of Surface Flinger faults.

For fault injection on the two native libraries (Bionic and SQLite), we performed 75 experiments (Figure~\ref{fig:benchmark_results_native_libs}), and we observed 36 failures for Samsung, 12 failures for Huawei, and 20 failures for HTC. In this case, the Samsung device has been the one that failed more frequently, but many failures also occurred in the other devices. In most of these failures, native processes such as the System Server failed because of unhandled exceptions and errors that were raised during filesystem I/O and during SQL queries.

We performed a close analysis of the experiments, in order to reproduce and to understand the causes of failures. In the following subsections, we present representative failures across the injected components.


\subsection{Failures in the Camera subsystem}
\label{sec:camera_detail_failures}

\begin{figure}[!htp]
    \centering

    \includegraphics[width=\columnwidth]{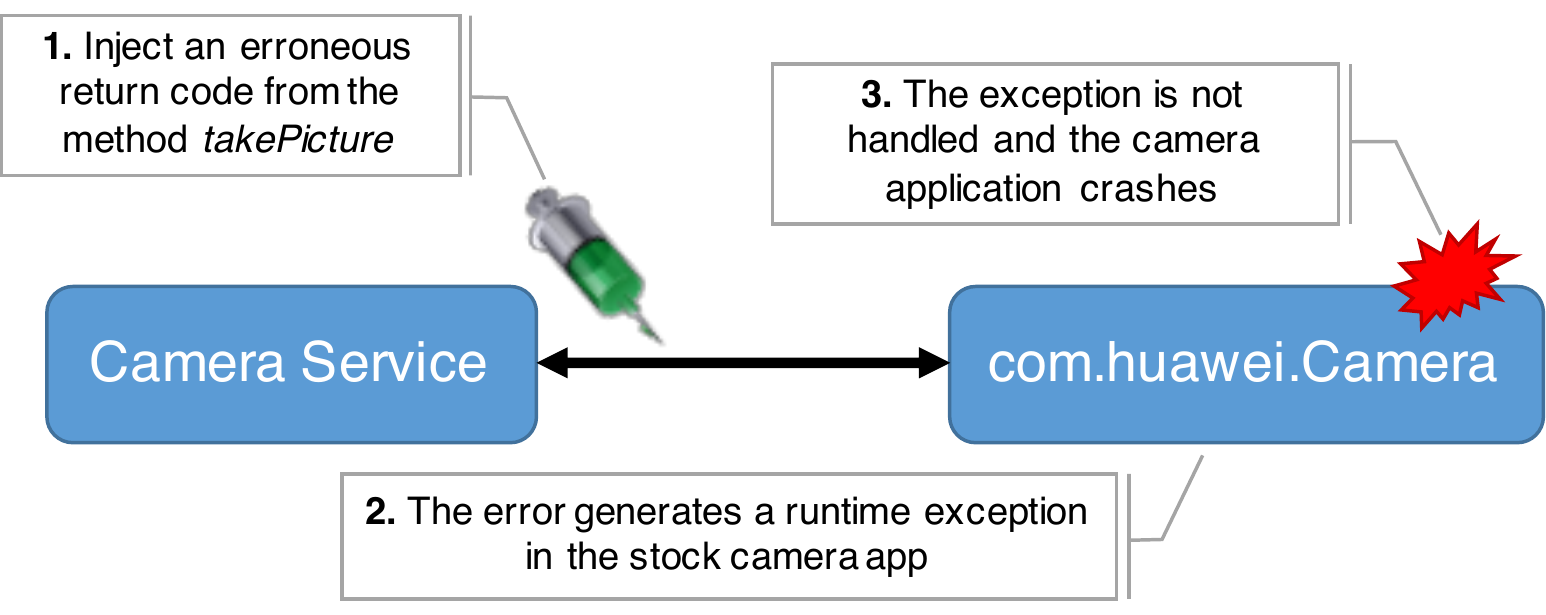}

    \vspace{-5pt}

    \caption{Failure of the Camera subsystem (faults in the Camera Service).}
    \label{fig:camera_failure_1}
\end{figure}

In the Camera subsystem, two relevant types of failures occurred during the experiments. 
The first type of failures was related to the Huawei Camera stock application, and caused by the injection of faults in the Camera Service. 
These experiments forced the Camera Service (such as, the \emph{takePicture} method) to return an error to the caller.
The effects of fault injection are showed in Figure~\ref{fig:camera_failure_1}. 
The error code returned by the method generates a run-time exception. This exception is not handled by the Camera stock application, thus the Camera application is aborted by the Android Run-Time. In this scenario, a black screen is showed to the user,  followed by a pop-up message that reports the process abort. This message does not provide any meaningful information to the user, and thus may give a bad perception of reliability.

\begin{figure}[!htp]
    \centering

    \includegraphics[width=\columnwidth]{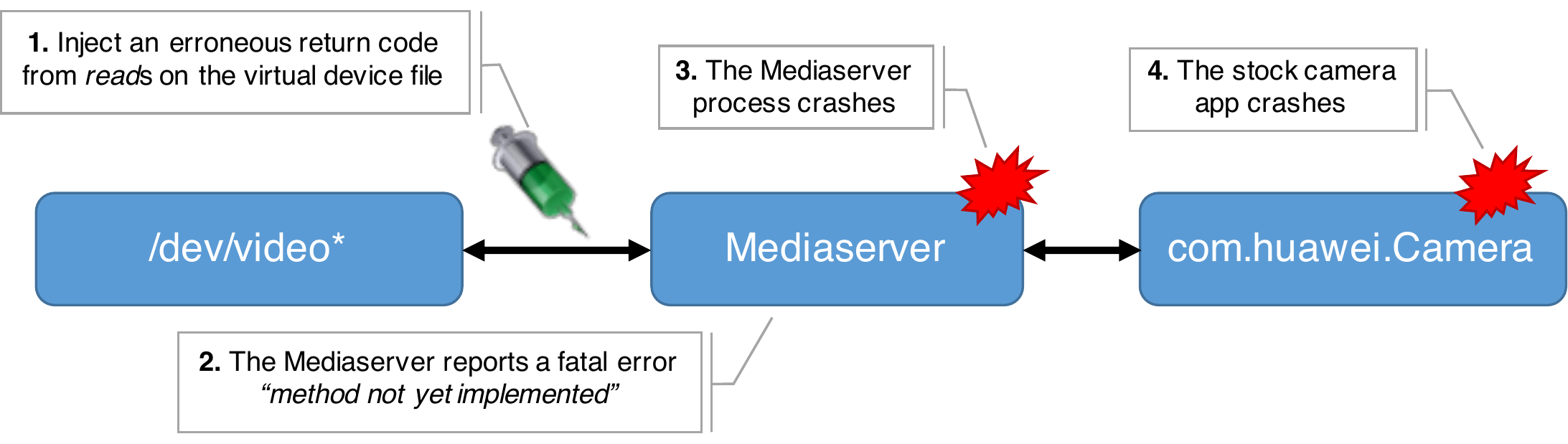}
 
    \vspace{-5pt}
 
    \caption{Failure of the Camera subsystem (faults in the Camera driver).}
    \label{fig:camera_failure_2}
\end{figure}

The second, more subtle, type of failure was caused by the injection of faults between the device driver and the Camera HAL in the Huawei device. In particular, faults were injected when the MediaServer process attempted to read from the \texttt{/dev/video} virtual device file, by forcing the operation to return an error code, such as \emph{ENOMEM} and \emph{ENODEV}. 
This scenario is showed in Figure~\ref{fig:camera_failure_2}. The injection led the MediaServer to fail with a crash. The MediaServer was not able to handle a corner case triggered by the fault injection: in the system logs, we found a fatal error message ``\emph{method not yet implemented}'' logged by the Camera HAL in the MediaServer. This error denotes that the vendor did not include an implementation of a method required by the Camera HAL API specification of the Android OS. This error was followed by the crash of the Media Server. In turn, the unavailability of the Media Server caused further exceptions towards the Huawei Camera stock application, which also crashed. 
The other devices handled the same faults in a more graceful way. 
In the HTC device, the Camera stock application catches the exception from the Camera Service. After the crash of the Media Server, both the Media Server and the Camera application are quickly restarted, without showing any error to the user. Thus, it is able to mask the fault to the user, and to provide a better perception of device reliability.

\vspace{2pt}
\noindent
\textbf{Improving reliability.} 
Fault injections pointed out that Android vendor customizations (such as the Camera HAL and stock app) are prone to missing or incomplete error handlers. In these scenarios, noticeable failure effects (black screens, cryptic error messages) are experienced by the end-user. Thus, it is advisable for vendors to mitigate these behaviors, such as by adding the missing exception handlers in the Camera stock application. This is confirmed by the analysis of the HTC device, in which the stock app is able to catch the exception, and to mask the fault through a soft restart of the Camera subsystem. Moreover, vendors should perform regression tests with fault injection to check  exception handling.


\subsection{Fault injection in the Phone subsystem}
\label{sec:phone_detail_failures}

\begin{figure}[!htp]
    \centering

    \includegraphics[width=\columnwidth]{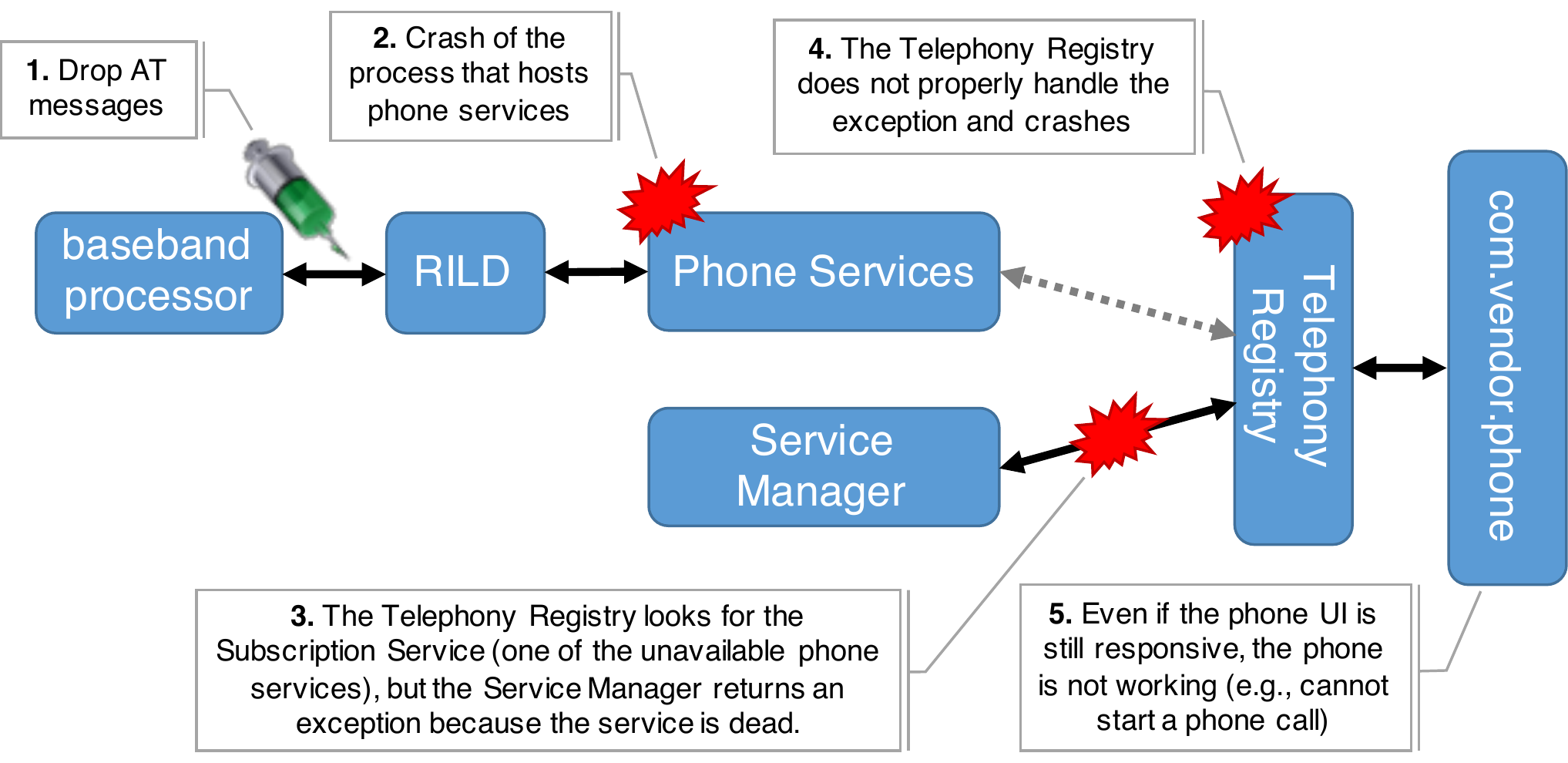}
    
    \vspace{-5pt}

    \caption{Failure of the Phone subsystem (faults in the modem/phone).}
    \label{fig:phone_subsystem_failure}
\end{figure}

This section describes an interesting scenario that involves the Phone subsystem, and in particular the RILD process, the Telephony Registry service, and the Huawei Phone stock application. 
In this scenario (see Figure~\ref{fig:phone_subsystem_failure}), faults are injected between the RILD and the baseband processor. The AndroFIT tool intercepts the AT messages flowing from the baseband processor to the RILD, and corrupts them by dropping the event codes and their parameters. These corruptions cause an incorrect internal state of the RILD, and cascade effects on phone services, such as \emph{isms}, \emph{phone\_huawei}, etc., leading to crashes. In turn, the Telephony Registry service crashes. Even if the phone services are automatically restarted, the device is not able anymore to manage events for the phone subsystem.

This failure impacts on the end-user, which is unable to perform phone calls. Even worse, the user is not informed about the problem, and the phone application becomes not responsive: when the phone stock application sends commands on behalf of the user, the commands are simply ignored by the phone subsystem, without showing any information regarding the unavailability of the phone subsystem.

\vspace{2pt}
\noindent
\textbf{Improving reliability.} 
This failure scenario involves several components, and points our several opportunities for improving reliability. 
The first, most important effect of the fault is the incorrect internal state of the RILD that causes the crash of phone services. Thus, it would be important for the RILD to recognize violations of the AT protocol, and gracefully handle them by adopting defensive programming practices, such as by checking at every step that the messages exchanged with the baseband processor follow the expected protocol. Moreover, the phone services should also be programmed defensively, by recognizing out-of-order events, and avoiding to crash in the case of these errors. 
Another opportunity of improvement is in the Huawei Phone stock application. It would be advisable to have mechanisms to detect that the phone subsystem is not responsive, for example by using a timeout when waiting for a response. Moreover, the application could trigger a soft restart to mask the error state and to retry the failed operation. The phone app should also provide some feedback to the user in these cases, since the user would have the perception of the lack of control over the device, and could get frustrated by the unsuccessful attempts to repeat the operation. Thus, in the case that these recovery mechanisms are not effective, the phone app should at least inform the user about the problem.


\subsection{Fault injection in the Sensors subsystem}
\label{sec:sensors_detail_failures}

\begin{figure}[htb!]
  \centering 
  \includegraphics[width=\columnwidth]{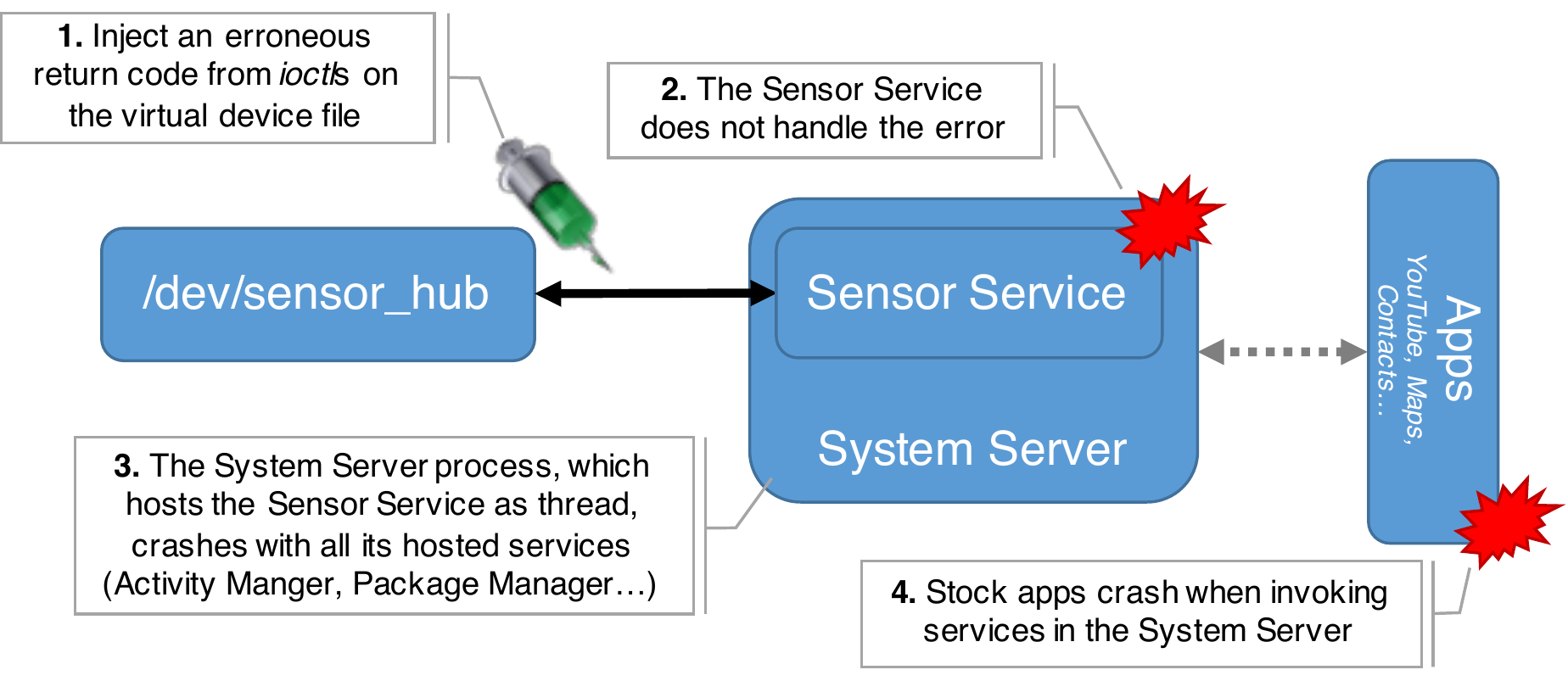}
  
  \vspace{-5pt}
  
  \caption{Failure of the System Server (faults in sensors' drivers).}
  \label{fig:sensors_failure}
\end{figure}

The Sensor subsystem exhibited the following severe failure behavior, that impacted not only on  the Sensors Service, but also on other subsystems of the Android OS. 
The failure scenario (see Figure~\ref{fig:sensors_failure}) is caused by faults injected when the Sensors Service accesses the sensor devices through virtual device files (e.g., \texttt{/dev/sensor\_hub}), such as returning \emph{ENOMEM} on I/O system calls.

The injections caused the crash of the Sensors Service. 
Since the Sensors Service executes within a thread of the System Server, the System Server process is also affected by the crash. In turn, this causes the termination of other Android services that execute inside the System Server. Most notably, the failure of the System Server affects the Package Manager. The Package Manager is a key service of the Android framework, since every access to privileged resources (such as files and hardware) has to be permitted by this service. Thus, the failure of the Package Manager causes cascading failures of the apps that require special permissions (e.g., Maps, Contacts, etc.).

\vspace{2pt}
\noindent
\textbf{Improving reliability.} 
This failure scenario is an example of failure propagation across different parts of the Android OS. In this case, the main weakness is in the co-existence of several services inside the System Server process. Thus, a fault in any service can potentially impact on all the other services. However, it is not simple to fix this problem since it is rooted in the design of the Android OS. Thus, for improving reliability, it is even more important to handle failures in these services, in order to prevent propagated failures of the whole System Server. In particular, the Sensors Service should check the successful outcome of I/O operations on the devices, and should gracefully handle any error to avoid crashes.


\subsection{Fault injection in the System Server}
\label{sec:sysserv_detail_failures}

\begin{figure*}[!htp]
    \centering
    
    \subfloat[\emph{stop\_activity} in the Activity Manager is delayed or stalled.\label{fig:sysserv_failure_2}]{
       \centering
       \includegraphics[width=0.18\textwidth]{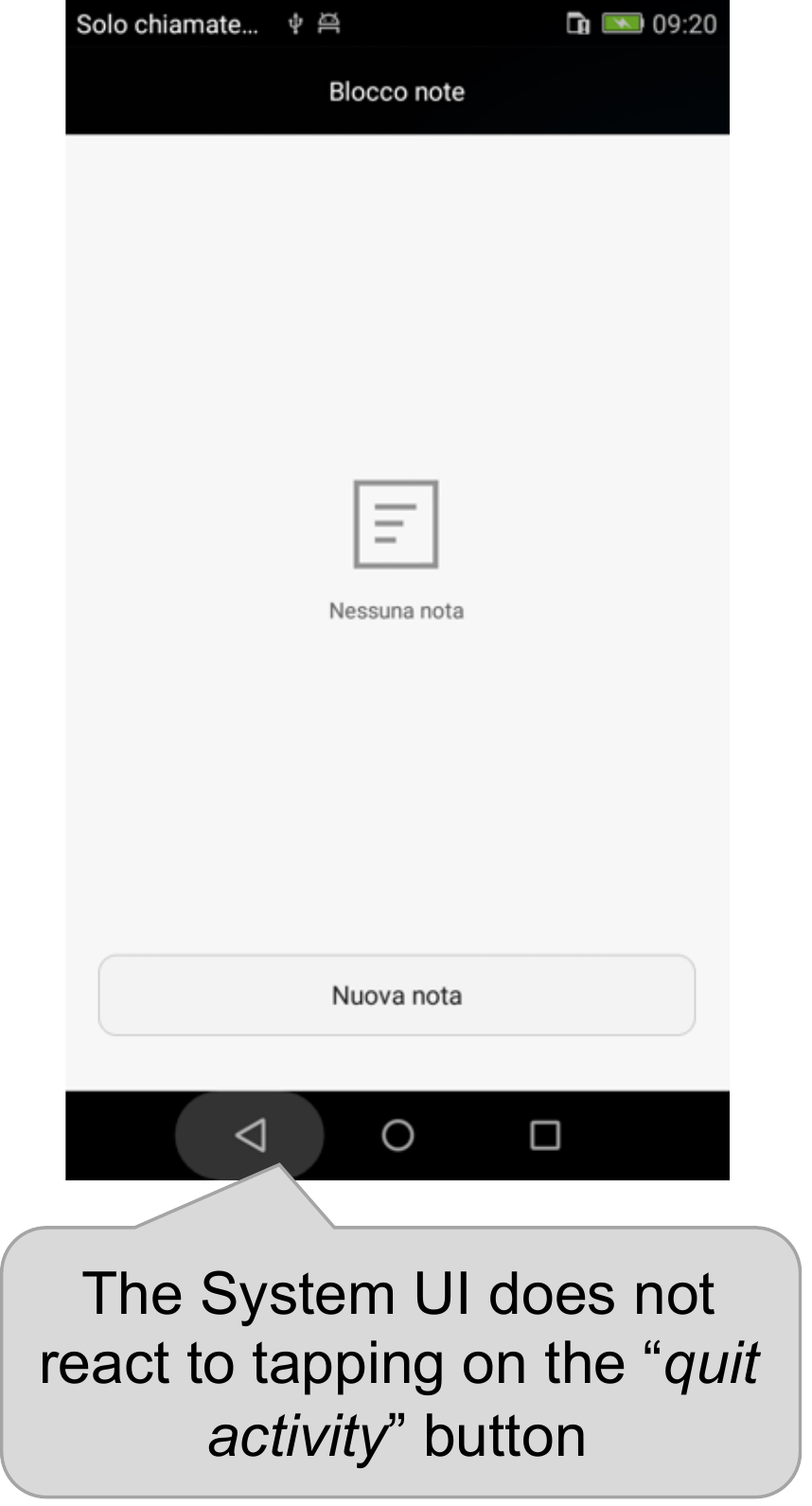}
    }
    \qquad\qquad
    \subfloat[\emph{bind\_service} in Activity Manager is delayed or stalled.\label{fig:sysserv_failure_3}]{
       \centering
        \includegraphics[width=0.27\textwidth]{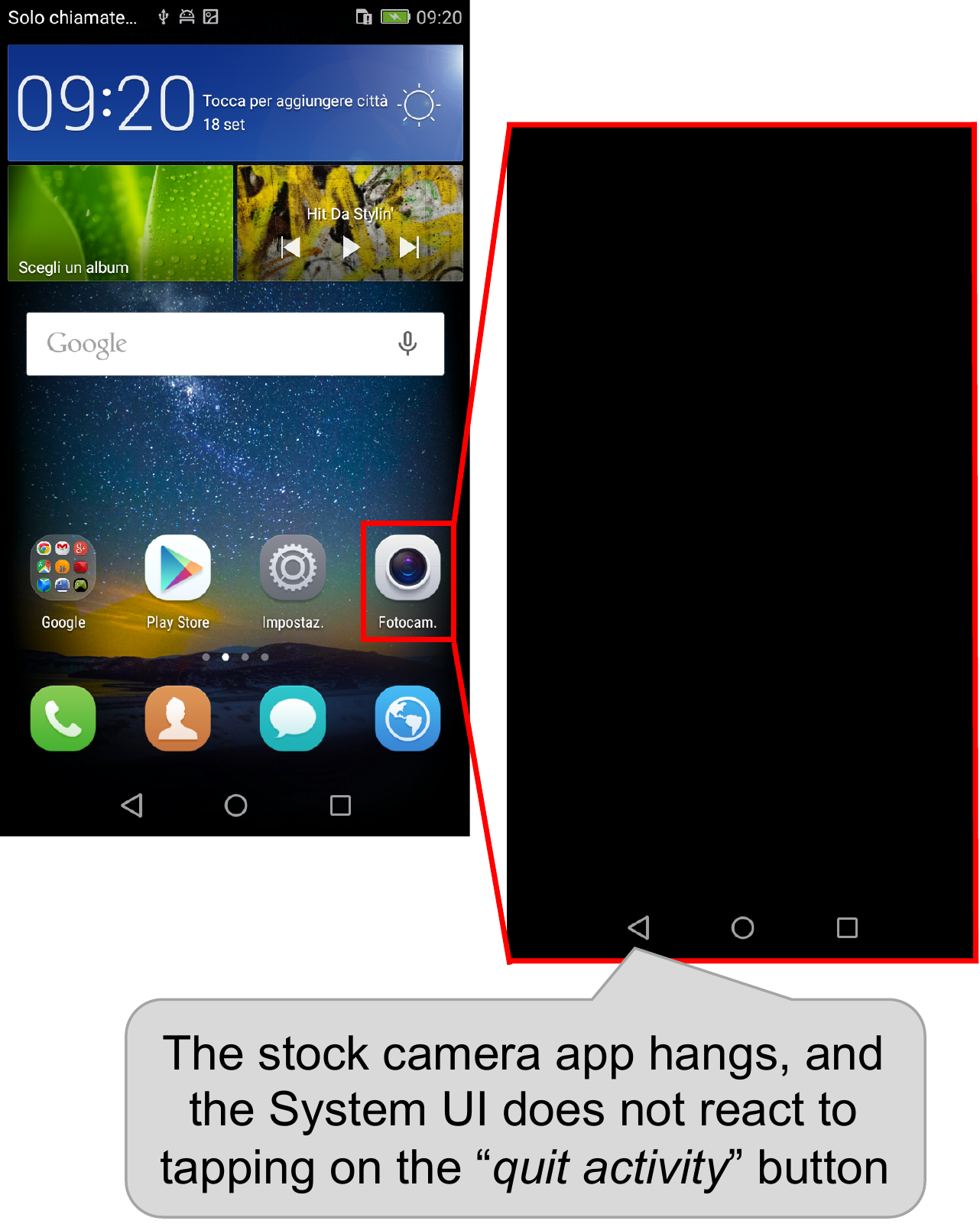}
    }
    \qquad\qquad
    \subfloat[\emph{resolve\_intent} in the Package Manager is delayed or stalled.\label{fig:sysserv_failure_4}]{
       \centering
       \quad
       \includegraphics[width=0.18\textwidth]{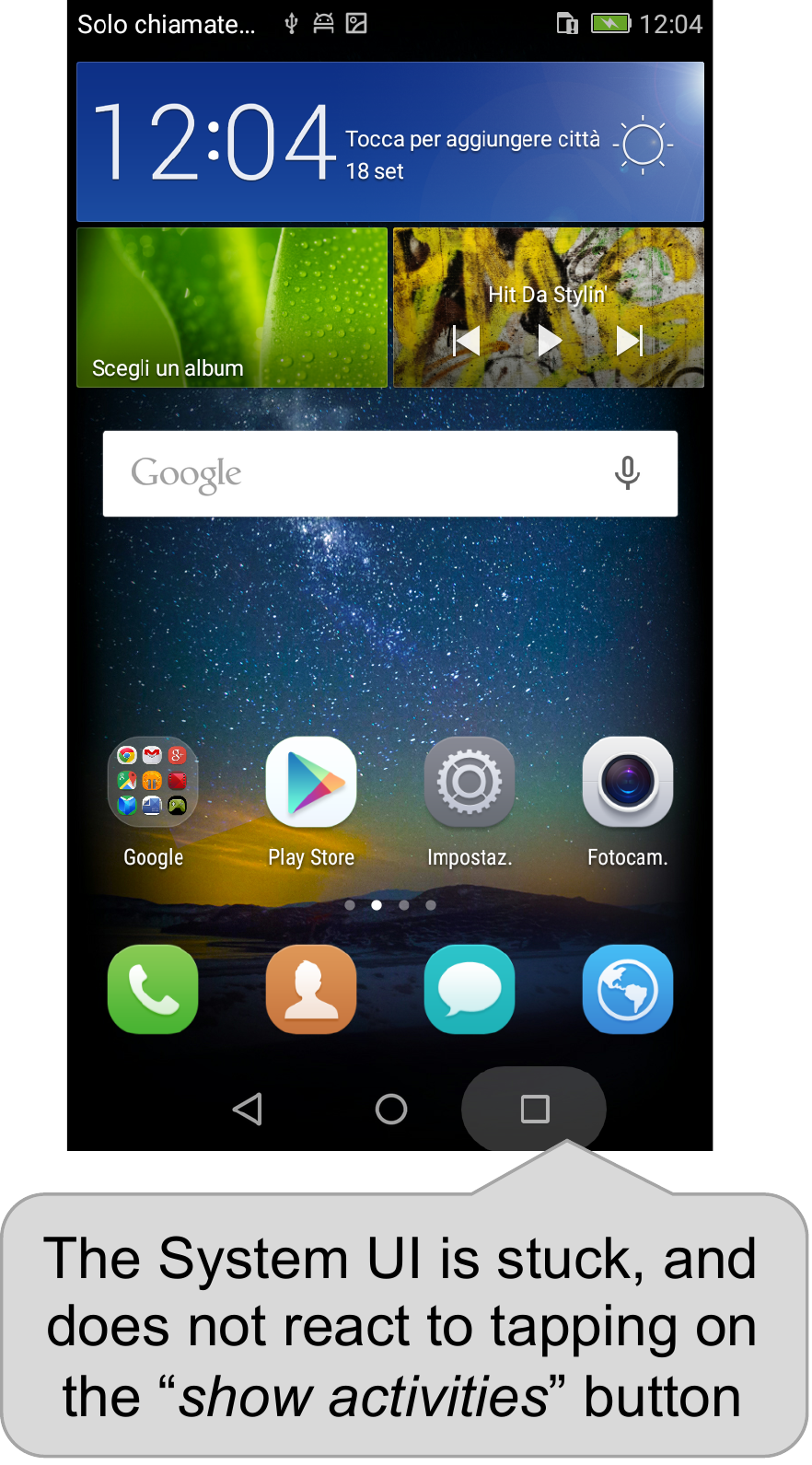}
       \quad
    }

    \caption{Impact on UI responsiveness caused by fault injection in the System Server.}
    \label{fig:phone_failure}
    
    \vspace{-10pt}
    
\end{figure*}

The AndroFIT tool tampered with invocations of the services offered by the System Server. These injections affected the consumers of these services, such as the System UI process, which handles the main elements (e.g., the notification bar, the main screen, etc.) of the user interface (UI), and starts apps, by invoking the Activity Manager and the Package Manager. 
One of these failure scenarios resulted in an unresponsive device, in which the user interface was stalled. These stalls are clearly noticed by the end-users, and negatively affect their perception of reliability. 
This failure occurred during the injection of a \emph{timeliness} fault in the \emph{stop\_activity} method of the Activity Manager service. The timeliness fault delays the execution of the \emph{stop\_activity} method by several seconds. This kind of fault occur when the device is overloaded (for example, due to a CPU hog app \cite{armando2012would,huang2015system}), or because of a performance bug in the System Server (e.g., due to software aging \cite{cotroneo2016software,qiao2018two}). 
When this timeliness fault is injected, the System UI process becomes not responsive (see Figure~\ref{fig:sysserv_failure_2}). If the user tries to leave the current activity (by tapping on the button in the bottom part the UI), the System UI invokes the \emph{stop\_activity} method, but it does not care whether the operation has been delayed or whether the current activity is still open. 
As a result, even if the user taps on the ``quit activity'' button several times, the System UI remains stuck. To avoid such undesirable behavior, the System UI should avoid getting stuck on the invocation, or enforce a timeout to detect the stall. The Android OS may attempt a recovery action, such as to force the termination of the activity by other means; or the system should inform the user that the operation is taking more time than expected. 
The injection also causes a restart of the System Server. If the user presses the ``show activities'' button, the Activity Manager will crash, and bringing down the whole System Server process.

Another failure scenario of stuck user interface involved both the Activity Manager and the Huawei Camera stock application. 
The AndroFIT tool injected a \emph{timeliness} fault on the invocation of the \emph{bind\_service} method of the Activity Manager. 
The \emph{bind\_service} method allows apps to setup long-running, background operations, and it is used by the Huawei Camera stock application to run a service when it is started for the first time. However, the call to \emph{bind\_service} can represent a bottleneck for the launch time of the Camera app. In our experiments, injecting a delay on \emph{bind\_service} causes the Camera application to become stuck with a black screen (see Figure~\ref{fig:sysserv_failure_3}). This behavior can be perceived as a severe failure by the users, since the launch time of basic apps is a key factor for the quality of experience \cite{tanenbaum2014modern,android-web-perf-anr}. 
This fault also caused a stall of the device UI, since even pressing the ``quit activity'' button does not allow the user to leave the Camera application and to perform other operations. Therefore, it is advisable to timely handle such stalls, as discussed for the previous failure.

Yet another case of stuck user interface occurred when injecting a \emph{timeliness} fault (again, a delay of several seconds), on the \emph{resolve\_intent} method offered by the Package Manager. This method is used by the system to manage broadcast requests for handling a file or event. In this case, the failure happened when the user presses the ``show activities'' button on the bottom part of the UI (see Figure~\ref{fig:sysserv_failure_4}). 
When the \emph{resolve\_intent} is injected with a delay, the whole System UI becomes unresponsive. The System UI does not show the list of current activities, and does not provide any feedback to the user. Even retrying to press the button does not solve the stall. Thus, it would be important to handle these UI stalls in order to avoid a poor user experience.

\vspace{2pt}
\noindent
\textbf{Improving reliability.} 
Since the System Server is a critical component, its faults often caused a stuck user interface. 
The stalls were caused by the fragile behavior of the System UI, which waits for a response for an indefinite amount of time, without enforcing a timeout. This is due to the fact that the System UI strictly relies on the responsiveness of the System Server. 
To prevent stalls, it is desirable, when possible, to have \emph{asynchronous} interactions with the System Server: that is, the System UI should not block waiting for a response (a \emph{synchronous} interaction), but it should be able to continue its execution, and to check whether the requested operation has actually been completed. However, asynchronous interactions are a more complex programming approach, and need to be carefully designed and tested. Thus, it is recommended to repeat the same tests with \emph{timeliness} faults to check that the new interactions are actually tolerant to delays. 

The Huawei Camera stock application is also affected by similar problems, since it can get stuck when it is started and the System Server is slow to respond. Since the quickness of the Camera application start is critical for the perceived responsiveness, it is important to optimize this use case, and to make it more robust. Again, these optimizations should be checked with the injection of \emph{timeliness} faults, in order to confirm that performance bottlenecks are avoided.


\subsection{Fault injection in native libraries}
\label{sec:native_libs}

Fault injection in the Bionic and SQLite native libraries pointed out potential crash failures of the System Server process. In the case of the Bionic library, we injected faults when the Package Manager used this library to read from storage APK files, in order to retrieve information on resources and permissions of the apps. 
The AndroFIT tool injected faults in I/O functions of the Bionic library (e.g., \emph{read}), by introducing corruptions in the contents of data buffers that are filled by these functions (e.g., by randomly changing few bytes of the data). 
The Package Manager crashed in the middle of the \emph{getPackageInfo} method, which raises an exception when it cannot retrieve and parse the data from a given application. The exception causes a crash of the whole System Server process in which the Package Manager service runs.

In the case of SQLite, we injected faults when the System Server operated on its internal database. We injected faults in several SQLite operations, including the opening of the database, the preparation of SQL queries, and the retrieval of SQL query results, by forcing these operations to return errors, to stall, to truncate data, etc.. 
For example, we observed crashes of the System Server process when its \emph{LockSettingsService} (i.e., the service that manages the lockscreen pattern or password, and related settings) stored and retrieved information about the device and the user. 
We found a crash failure of this service, and of the System Server as a whole, due to unhandled exceptions. We injected an \emph{unavailability} fault in the \emph{sqlite\_step} operation of the SQLite library (i.e., when retrieving the tuples generated by an SQL query) by forcing an error code (\emph{SQLITE\_ERROR}). In turn, the JNI wrapper around the SQLite library throws an unhandled exception.

\vspace{2pt}
\noindent
\textbf{Improving reliability.} 
The Package Manager needs to isolate the effects of corrupted APK files, by only affecting the application for which metadata could not be retrieved. Thus, the app should be aborted, or should not be started at all, without affecting the System Server and other applications. This would require to carefully check that the contents read from the APK are not corrupted, by performing checks that the data are reasonable (for example, by checking that strings have invalid characters or are too long, or checking that integer variables should have values within a range, etc.). These defensive checks should be then tested by means of fault injection in the contents of the APK files.

In the case of SQL queries, the System Server and the stock apps should catch any exception that might occur, and they should avoid the crash by masking the exception. In the specific case of the lock settings, the device should inform the user that lockscreen information could not be retrieved due to database errors. The device could offer the user an alternative way to unlock the device (for example, asking for a different PIN or password). Another approach could be use store and reuse a previous version of the database in the case of problems. In the case that none of these alternatives is feasible, the device should not block the device but it should let in the user, since the database error does not allow to correctly enforce the protection. Most importantly, the system should not crash in the case of database errors.


\subsection{Fault injection in Surface Flinger}
\label{sec:surface_flinger}

We observed UI failures when injecting faults on the Binder service API of the Surface Flinger. In particular, we inject faults (delays, service refused, data corruptions) on APIs for creating a new connection to the Surface Flinger (\emph{createConnection}, \emph{createDisplayEventConnection}), and to update metadata about surfaces that are produced (\emph{setTransactionState}). These APIs are used by system processes and services, including the Window Manager, the System UI, and the Launcher, thus they can have  a severe impact on the UI.

When the Surface Flinger is invoked by the \emph{RenderThread} task inside one of these components, the UI freezes for several seconds, and the process experiences a crash. Then, the Launcher and SystemUI processes are automatically restarted, but the problem still persists; this leads to repeated restarts of the processes (up to 25 times in a row in our experiments). Instead, the Android OS does not perform any recovery action on the actual root cause of the problem (i.e., the Surface Flinger, in which we injected a fault). The device was recovered only by physically forcing a reboot of the device.

The injected faults also impacted on the shared memory buffers that are used to drive the graphics hardware through HAL APIs. Examples of high-severity error messages include: ``\emph{queueBuffer: error queuing buffer to SurfaceTexture}'', ``\emph{dequeueBuffer: can't dequeue multiple buffers without setting the buffer count}'', and ``\emph{Channel is unrecoverably broken and will be disposed}'' in the InputDispatcher. These errors were often followed by the forced termination of proprietary services (e.g., \emph{com.huawei.systemmanager.rainbow.service}).

\vspace{2pt}
\noindent
\textbf{Improving reliability.} 
Fault injection in the Surface Flinger revealed that the Android OS is sensitive to faults in this process, since several other processes may stall or crash. The impact of the faults also involve system processes such as the Window Manager, the Launcher, and the System UI. 
The recovery mechanisms in the Android OS were not sufficient to address these faults, since they acted on the system processes were the fault propagated, rather than the root cause. 
Therefore, the Surface Flinger is neglected by fault recovery. Instead, the Android OS should also address the Surface Flinger, by restarting it or by freeing resources in order to allow the Surface Flinger to recover from the fault. 
Moreover, the Android OS needs exception handlers that are able to correctly handle faults in this component.

\subsection{Discussion}

We looked in retrospective at which injected components propagated failures across the Android OS, and which types of injections caused these failures. 
In \tablename{}~\ref{fig:fault_model_discussion}, we list the types of injections performed by the tool (\emph{unavailability}, \emph{timeliness}, \emph{corruption}, \emph{resource management}) and, for each type, we denote with marks which Android subsystems were vulnerable to the type (i.e., more than 20\% of the injections caused a failure of system). 
According to the experimental results, we can draw the following lessons learned.

\begin{table}[!htb]
  \centering
  
  \vspace{-5pt}

  \caption{Overview of components and fault types that caused failures.}
  \label{fig:fault_model_discussion}
 
  \includegraphics[width=\columnwidth]{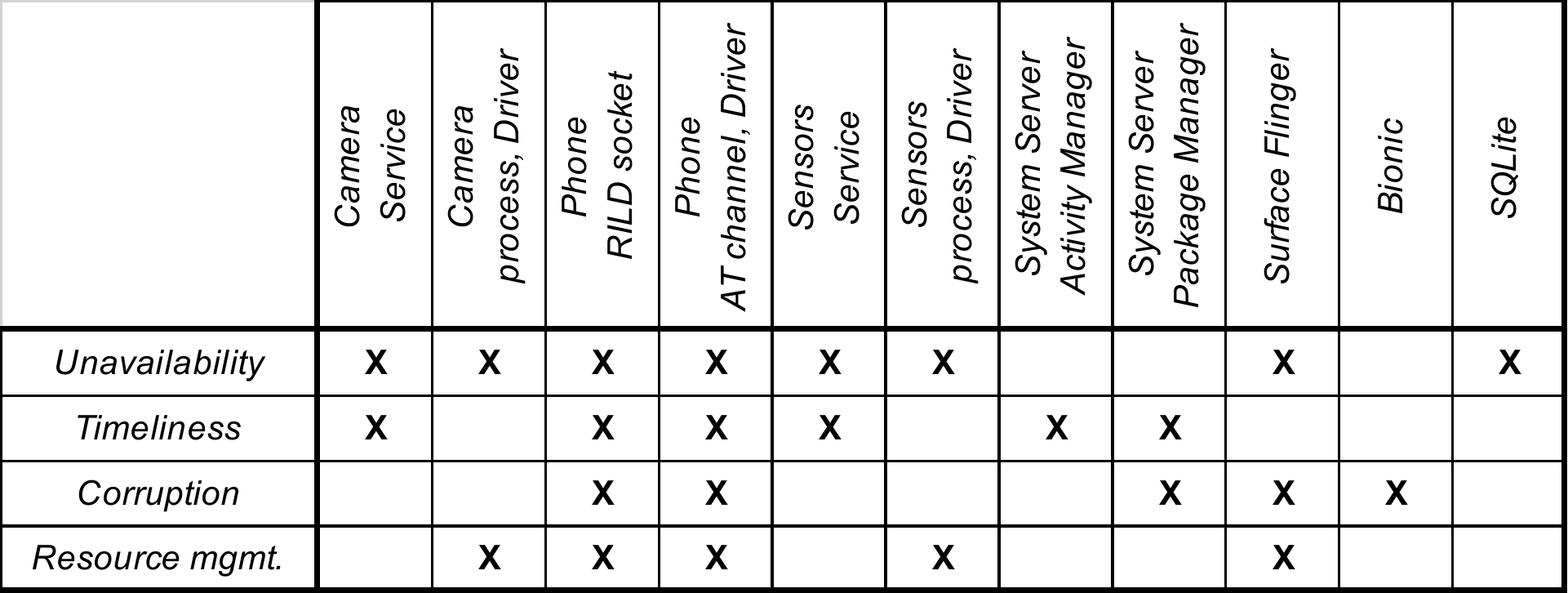}
  
  \vspace{-5pt}
  
\end{table}

The \textbf{unavailability} faults (that is, exceptions and error codes returned by APIs, such as Binder calls, library calls and system calls) were the ones that found issues in the highest number of Android subsystems. In these vulnerabilities, the Android OS lacked exceptions or errors handlers, thus the exception/error was able to spread and cause the failure of Android services and applications. Since so many Android subsystems were vulnerable to these faults, \textbf{it is recommended to always include this fault type in fault injection test plans}. Another advantage of this fault type is that it can provide clear and easy suggestions for improving reliability: they point out the specific exceptions/errors that are not tolerated, thus the developers can mitigate them by implementing the missing exception/error handlers. This is especially important in the stock applications, as they must provide user-friendly feedback in the case of faults, in order to give a good perception of the reliability of the device. For example, we found cases in which the Huawei stock camera app did not manage the exceptions generated from the Camera Service.

The \textbf{timeliness} faults (that is, delays and stalls of API calls) were another frequent cause of failures of the Android OS. In particular, when the delay/stall occurs in four Android services (the Camera Service, the Sensors Service, the Package Manager and the Activity Manager), these services cause the failure of stock apps and of the System UI. This behavior happens when the application invokes the service in a \emph{synchronous} way (that is, the application stops until the service is provided). Unfortunately, the \emph{synchronous} approach is a cause of performance bottlenecks for the application, and it can cause failures if the API is delayed or stalled. The experiments show that \textbf{timeliness faults have a severe impact when injected on the Binder APIs of Android services}, since stock applications are often vulnerable to this type of faults. Moreover, the delays/stalls of UI applications must be avoided since are clearly noticed by the user, and would cause a poor quality of experience. In order to make the apps more robust against these faults, they should either adopt an \emph{asynchronous} approach to call the service (by allowing the app to continue to be responsive even if the call is delayed/stalled); or the apps should enforce a timeout to detect the long execution time of the service, and retrying the operation, or aborting the operation with a user-friendly notification.

The \textbf{corruption} faults were effective against specific components (the RILD socket and AT channel, the Surface Flinger and the Bionic library) that handled structured data, causing the crash of key services: 
AT channel corruptions (e.g., a correct AT command is dropped or replaced with a wrong one) crashed the RILD; 
Surface Flinger corruptions (e.g., wrong transaction state of the streamed surfaces) crashed the SystemUI; 
filesystem I/O corruptions (e.g., APK metadata) through Bionic crashed the Package Manager. 
These results point out that the corruption of \emph{protocols} (such as the AT protocol) and \emph{formats} (such as the APK format, and the transaction format in surface streams) can expose the Android OS to failures. Indeed, it is difficult for developers to build robust protocol/format parsers that could manage any invalid data in the protocol/format. Therefore, we recommend that \textbf{corruptions should be injected into protocols (such as the AT protocol) and formats (such as APK metadata) that are complex and tricky to parse/handle robustly}. We found that even a simple approach (such as injecting random noise in these protocols/format) can be effective to highlight vulnerabilities. 
Instead, we found that other components (such as the Camera Service, Sensors Service, Activity Manager, etc.) are quite insensitive to corruptions, since these services do not expose complex protocols/formats. In these cases, the injection corrupted the input/output parameters of the services (for example: in the Camera Service, parameters such as \emph{whitebalance=auto} are replaced with incorrect values, and numeric values are corrupted with 0, negative, MAX, or random values; in the Activity Manager, the methods return Intents with an incorrect Action field, such as ACTION\_BATTERY\_CHANGED replaced by ACTION\_POWER\_CONNECTED, or a truncated Data URI). In other cases, such as SQLite, the corruptions caused the SQL query results to be truncated. These injections can affect individual applications by corrupting their output (for example, the Camera application can return distorted images, or a background app service may not be loaded); but these injections do not affect the stability of the Android OS and stock apps (neither fatal exceptions nor ANRs occurred).

The \textbf{resource management} faults (such as, the exhaustion of memory, the inability to open files or create threads, etc.) were effective to find vulnerabilities in processes and components in the \emph{native layer}. Since these parts are written in C/C++, they do not benefit from robust and automated resource management (as it would be the cause for the Java language), and thus they are often vulnerable to resource-related problems. Thus, we advise to \textbf{inject resource management faults for testing the robustness of components and processes in the native layer}. Examples of this are the RILD process and the Media Server (which runs the Camera Service), as we found that these processes were affected by failures in the case of resource unavailability.

\section{Limitations and future directions}
\label{sec:limitations}

The problem of emulating software faults is a very difficult one, and the proposed approach is not a definitive solution to this problem. Indeed, one important open challenge is that the enumeration of failure modes can be still incomplete, since the system can also exhibit more complex failure behaviors, which are not limited to an individual service call and which span across a sequence of service calls. If these complex behaviors are included in the fault model space, then the enumeration can become exceedingly large due to combinatorial explosion. On the one hand, this makes the search space too large to be exhaustively covered. On the other hand, the SIR approach provides an upper bound (even if large) to the search space, as the number of service interfaces and resources is finite, while the space of "faulty programs" can be infinite. Thus, focusing on service interfaces and resources helps test engineers by allowing them to sample the failure space according to their testing budget (e.g., the amount of time that they can spend on performing fault injection tests). For example, one possible sampling policy could be to cover in depth single-service failures, and to sample multiple-service failures (from $2$- up to $k$-way combinations) with gradually lower sampling density.

In general, selecting an efficient, yet realistic, fault model to inject is the main challenge in the field of software fault injection. For example, in the case of value failures, in which one or more output data of the system can take incorrect values from the data domain, it is difficult to select faulty values from the domain. Moreover, it is difficult to anticipate where in the software the failures can occur, such as which parts of the software can be subject to a timeliness failure (e.g., non-terminating loops, an incorrect transition in a state machine). In the SIR approach, the failure modes are derived from qualitative guidelines, which based on evidence from previous studies.

In our ongoing work, we have been studying empirical evidence for a more systematic selection of the fault model, by analyzing the effects of injected bugs on software interfaces \cite{natella2018analyzing}. For example, the empirical data showed pointed out which value failures are the most common ones (e.g., boundary values such as NULL pointers, and values differing from the correct ones by a small offset); that the failures are concentrated in one or few service calls, expect in the case of API meant to be called in loops; and other empirical findings about the extent of data corruptions, the occurrence of failure signals such as API error codes, etc.

Future work is needed to link these empirical findings with an automated fault injection tool, in order to assure the realism of the injected faults, and to reduce the manual work needed to create a fault model. For example, as suggested by the reviewer, the SIR approach could leverage static analysis to identify service interfaces that are prone to the failure modes, by looking at the complexity of method implementations behind the service interfaces, such as the cyclomatic complexity of methods, the depth of the method call graph, the number and type of dependencies on external services. This information can be adopted to perform more efficient and realistic injections, by avoiding to inject in trivial parts of the system that are unlikely to experience failures.

Finally, we remark that a complete assessment of mobile devices should also encompass security vulnerabilities, since an insecure system with an attack surface exposed to threats cannot be considered reliable. These security vulnerabilities are not addressed by the proposed approach, as it focuses on assessing the impact of accidental, non-malicious faults on the mobile user experience.

\section{Related work}
\label{sec:related_work}

We discuss studies on robustness testing of mobile systems, including fuzz testing, GUI testing, and fault injection.

\vspace{1pt}
\noindent
\textbf{Fuzz Testing.} 
Fuzzing is a technique that generates large volumes of random data to test complex software interfaces. 
In the context of mobile devices, Miller et al. \cite{miller2012ios} presented the \emph{zzuf} fuzzing tool for iOS apps, which intercepts and randomly mutates multimedia input files. The tool was effective for testing media players, image viewers, and web browsers, because of the quantity and complexity of inputs for these programs. Lee et al. \cite{lee2015designing} designed the \emph{Mobile Vulnerability Discovery Pipeline} (MVDP), an approach that generates random, invalid multimedia input files to crash Android and iOS apps. They developed heuristics to increase the efficiency of fuzzing, and strategies for scaling large numbers of fuzz tests on farms of smartphones. In addition to multimedia data, Android fuzzers have been adopted to attack network and IPC interfaces. Mulliner et al.  \cite{mulliner2009fuzzing} found severe vulnerabilities in the SMS protocol. \emph{Droidfuzzer} \cite{ye2013droidfuzzer} and \emph{Intent Fuzzer} \cite{sasnauskas2014intent} targeted Android activities that accept data through Android Intents. \emph{Chizpurfle} \cite{iannillo2017chizpurfle} and \emph{BinderCracker} \cite{feng2016bindercracker} performed fuzz testing on Android OS services. We remark that all these fuzzing solutions are meant to test input validation of individual components (apps and services). This work is complementary to fuzz testing, as it evaluates how the mobile OS as a whole behaves \emph{after} that an individual component fails.

\vspace{1pt}
\noindent
\textbf{App Testing.} Most of the research on mobile testing focuses on individual mobile apps, by generating input UI events and data. The \emph{Monkey} \cite{android-monkey} is a well-known UI exerciser tool from the Android project, which randomly generates user events such as clicks, touches, or gestures. \emph{Dynodroid} \cite{machiry2013dynodroid} extends Monkey by extracting system events from the apps, and instrumenting the Android framework to guide UI event generation. \emph{EvoDroid} \cite{mahmood2014evodroid} uses an evolutionary algorithm, enhanced by a static analysis of the apps to identify where to apply crossover and mutation of the test sequences. \emph{T+} \cite{linares2015enabling} records UI events under normal application usage, and re-arranges these events to generate new test cases. Adamsen et al. \cite{adamsen2015systematic} and Zhang and Elbaum \cite{zhang2014amplifying} amplified existing GUI test suites, by introducing exceptions to expose test cases to adverse conditions, \ie unexpected events that may interfere with the execution of the test. The tests are executed in parallel with multiple Android emulator instances, and the Android framework is instrumented to control the execution of a test and to perform event injections. \emph{Caiipa} \cite{liang2014caiipa} follows a similar approach, by providing a cloud service for testing Windows mobile apps. The apps are stressed with random GUI events under several ``contexts'', i.e., unexpected conditions (\eg network connectivity and availability of sensors), distributing the tests among both emulators and actual devices.

\vspace{1pt}
\noindent
\textbf{Fault Injection.} The previous approaches test individual components with respect to external inputs, events or conditions; instead, fault injection evaluates a system with respect to faults that originate from components inside the system. \emph{G-SWFIT} \cite{duraes2006:emulationswfaults} is a technique that emulates software faults in \emph{off-the-shelf} components, by mutating their binary code. This approach has been adopted for benchmarking the dependability of systems, such as different releases of the Windows OS with respect to faulty device drivers \cite{duraes2002characterization} and different web servers with respect to faults in the underlying OS \cite{duraes2004:generic-faultloads}.  \emph{PAIN} \cite{winter2015no} applied this approach to the Android OS, by injecting software faults into its device drivers, and by parallelizing tests to scale them. However, PAIN was limited to faults in the OS kernel, and did not consider the general case of faulty software components in the wider architecture of the Android OS. The Android OS includes dozens of components, which provide abstractions and services in several areas such as the connectivity (e.g., Bluetooth, WiFi, phone), multimedia (e.g., camera, audio), sensors (e.g., GPS), etc.. Our work expands the scope of the analysis to the whole Android OS, and leverages a lightweight fault modeling approach to define faults for these components.

\section{Conclusion}
\label{sec:conclusion}

In this work, we proposed an approach for designing and executing fault injection experiments on the Android OS. We applied the approach to systematically define a large set of faults across the components on the Android system. Then, we implemented these faults into an automated fault injection tool. Finally, we performed fault injection experiments on three commercial Android devices. We found reliability issues across all of these devices: failures were due both to weak spots in the general Android architecture (such as, cascading failures of services running in the same process), and to customizations of the Android OS by the vendors (such as, the System UI, vendor HALs, and stock apps).

\section*{Acknowledgments}
This work has been partially supported by UniNA and Compagnia di San Paolo in the frame of Programme STAR, and by Huawei Technologies Co., Ltd.

\ifCLASSOPTIONcaptionsoff
  \newpage
\fi



%

\bibliographystyle{IEEEtran}
\bibliography{IEEEabrv,references,references-thesis,references-android}

%


\begin{IEEEbiographynophoto}{Domenico Cotroneo} (Ph.D.) is associate professor at the Federico II University of Naples. His research interests include software fault injection, dependability assessment, and field-based measurements techniques.
\end{IEEEbiographynophoto}

\begin{IEEEbiographynophoto}{Antonio Ken Iannillo} (Ph.D.) is research associate at the Interdisciplinary Centre for Security, Reliability, and Trust (SnT), University of Luxembourg. His research activities focus on software dependability and security assessment.
\end{IEEEbiographynophoto}

\begin{IEEEbiographynophoto}{Roberto Natella} (Ph.D.) is assistant professor at the Federico II University of Naples, Italy. His research interests include dependability benchmarking, software fault injection, software aging and rejuvenation, and their application in OS and virtualization technologies.
\end{IEEEbiographynophoto}

\begin{IEEEbiographynophoto}{Stefano Rosiello}(Ph.D.) is post-doc researcher at the Federico II University of Naples. 
His research interests include overload control, experimental reliability evaluation, and dependability benchmarking, focusing on network function virtualization and cloud infrastructures. 
\end{IEEEbiographynophoto}




\end{document}